\documentclass[preprint]{aastex}
\usepackage{graphicx}
\usepackage{longtable}
\usepackage{textcmds}
\usepackage{endfloat}
\usepackage{amsmath}
\usepackage{dcolumn}
\newcolumntype{.}{D{.}{.}{-1}}

\shorttitle{Orbit and Bulk Density of (101955) Bennu}
\shortauthors{Chesley et al.}
\slugcomment{Accepted by Icarus on February 19, 2014}

\begin{document}

\title{Orbit and Bulk Density of the OSIRIS-REx Target Asteroid (101955) Bennu}

\author{
Steven R. Chesley\altaffilmark{*}\altaffilmark{1}, 
Davide Farnocchia\altaffilmark{1},
Michael C. Nolan\altaffilmark{2}, 
David Vokrouhlick\'y\altaffilmark{3},
Paul W. Chodas\altaffilmark{1},
Andrea Milani\altaffilmark{4},
Federica Spoto\altaffilmark{4},
Benjamin Rozitis\altaffilmark{5},
Lance A. M. Benner\altaffilmark{1},
William F. Bottke\altaffilmark{6},
Michael W. Busch\altaffilmark{7},
Joshua P. Emery\altaffilmark{8},
Ellen S. Howell\altaffilmark{2},
Dante S. Lauretta\altaffilmark{9},
Jean-Luc Margot\altaffilmark{10},
and
Patrick A. Taylor\altaffilmark{2}
}

\altaffiltext{*}{\tt steve.chesley@jpl.nasa.gov}

\altaffiltext{1}{Jet Propulsion Laboratory, California Institute of
Technology, 4800 Oak Grove Drive, Pasadena, CA 91109, USA}

\altaffiltext{2}{Arecibo Observatory, Arecibo, PR, USA}

\altaffiltext{3}{Charles Univ., Prague, Czech Republic}

\altaffiltext{4}{Univ. di Pisa, Pisa, Italy}

\altaffiltext{5}{Open Univ., Milton Keynes, UK}

\altaffiltext{6}{Southwest Research Institute, Boulder, CO, USA}

\altaffiltext{7}{SETI Inst., Mountain View, CA, USA}

\altaffiltext{8}{Univ. Tennessee, Knoxville, TN, USA}

\altaffiltext{9}{Univ. Arizona, Tucson, AZ, USA}

\altaffiltext{10}{Univ. California, Los Angeles, CA, USA}



\begin{abstract}
The target asteroid of the OSIRIS-REx asteroid sample return mission, (101955) Bennu (formerly 1999~RQ$_{36}$), is a half-kilometer near-Earth asteroid with an extraordinarily well constrained orbit. An extensive data set of optical astrometry from 1999--2013 and high-quality radar delay measurements to Bennu in 1999, 2005, and 2011 reveal the action of the Yarkovsky effect, with a mean semimajor axis drift rate $da/dt = (-19.0 \pm 0.1)\times 10^{-4}$ au/Myr or $284\pm 1.5\;\rm{m/yr}$. The accuracy of this result depends critically on the fidelity of the observational and dynamical model. As an example, neglecting the relativistic perturbations of the Earth during close approaches affects the orbit with $3\sigma$ significance in $da/dt$. 

The orbital deviations from purely gravitational dynamics allow us to deduce the acceleration of the Yarkovsky effect, while the known physical characterization of Bennu allows us to independently model the force due to thermal emissions. The combination of these two analyses yields a bulk density of $\rho = 1260\pm70\,\rm{kg/m^3}$, which indicates a macroporosity in the range $40\pm10$\% for the bulk densities of likely analog meteorites, suggesting a rubble-pile internal structure. The associated mass estimate is $(7.8\pm0.9)\times 10^{10}\, \rm{kg}$ and $GM = 5.2\pm0.6\,\rm{m^3/s^2}$.

Bennu's Earth close approaches are deterministic over the interval 1654--2135, beyond which the predictions are statistical in nature. In particular, the 2135 close approach is likely within the lunar distance and leads to strong scattering and therefore numerous potential impacts in subsequent years, from 2175--2196. The highest individual impact probability is $9.5\times 10^{-5}$ in 2196, and the cumulative impact probability is $3.7\times 10^{-4}$, leading to a cumulative Palermo Scale of -1.70.
\end{abstract}


\section{Introduction}\label{sec:intro}

The Apollo asteroid (101955) Bennu, a half-kilometer near-Earth asteroid previously designated 1999~RQ$_{36}$, is the target of the OSIRIS-REx sample return mission. A prime objective of the mission is to measure the Yarkovsky effect on this asteroid and constrain the properties that contribute to this effect. This objective is satisfied both by direct measurement of the acceleration imparted by anisotropic emission of thermal radiation, the first results of which are reported here, and by constructing a global thermophysical model of the asteroid to confirm the underlying principles that give rise to this effect.

Bennu was discovered by the LINEAR asteroid survey in September 1999. Since then, more than 500 optical observations have been obtained for this Potentially Hazardous Asteroid (PHA).  Moreover, the asteroid was observed using radar by the Arecibo and Goldstone radio telescopes during three different apparitions. Thanks to this rich observational data set, Bennu has one of the most precise orbits in the catalog of known near-Earth asteroids. The exceptional precision of the Bennu orbit allows one to push the horizon for predicting possible Earth impacts beyond the 100 years typically used for impact monitoring \citep{milani_etal_nonlinear}, and indeed \citet{milani_etal_rq36} showed that Earth impacts for Bennu are possible in the second half of the next century. In particular, the cumulative impact probability they found was approximately $10^{-3}$, about half of which was associated with a possible impact in 2182. However, the occurrence of an impact depends decisively on the Yarkovsky effect because the prediction uncertainty due to this nongravitational perturbations dominates over the orbital uncertainty associated with astrometric errors.

The Yarkovsky effect is a subtle nongravitational perturbation that primarily acts as a secular variation in semimajor axis and thus causes a runoff in orbital anomaly that accumulates quadratically with time \citep{bottke_etal06}. The computation of the Yarkovsky perturbation requires a rather complete physical model of the asteroid, including size, shape, density, spin rate and orientation, thermal properties, and even surface roughness \citep{rozitis_green_yarko}. Though such a complete profile is rarely available, the orbital drift due to the Yarkovsky effect can sometimes be determined from an asteroid observational data set. For example, \citet{chesley_golevka} managed to directly estimate the Yarkovsky effect for asteroid (6489) Golevka by using three radar apparitions. \citet{1992bf} employ the Yarkovsky effect to match precovery observations of asteroid (152563) 1992~BF that are incompatible with purely gravitational dynamics. More recently \citet{nugent_etal_2012} and \citet{farnocchia_yarkolist_2013} have estimated the Yarkovsky effect for a few tens of near-Earth asteroids by using a formulation that depends on a single parameter to be determined from the orbital fit.

Besides Bennu, there are two other asteroids for which possible impacts are known to be driven by the Yarkovsky effect: (29075) 1950~DA \citep{giorgini_1950da} and (99942) Apophis \citep{chesley_acm05, giorgini_apophis_08}. The relevance of the Yarkovsky effect for Apophis is due to a scattering close approach in 2029 with minimum geocentric distance $\sim$38000 km. For 1950 DA the influence of the Yarkovsky effect for an impact in 2880 is due to the long time interval preceding the potential impact. However, no estimate of the Yarkovsky perturbation acting on these two asteroids is currently available. To analyze such cases one can use the available physical constraints for the specific objects, along with general properties of near-Earth asteroids (e.g., albedo, thermal inertia, bulk density, etc.) to statistically model the Yarkovsky effect. The orbital predictions and the impact hazard assessment are then performed by a Monte Carlo simulation that accounts for both the Yarkovsky effect distribution and the orbital uncertainty \citep{farnocchia_etal_apophis, farnocchia_etal_1950da}. For Bennu, no such heroics are required. As we shall see, we now have a precise estimate of the orbital deviations caused by the Yarkovsky effect, as well as a comprehensive physical model distilled from numerous investigations.

While the Yarkovsky effect requires a priori knowledge of several physical parameters to be computed directly, its detection through orbital deviations can be used to constrain the otherwise unknown physical parameters. When the spin state is unknown, one can derive weak constraints on obliquity, as was first shown by \citet{1992bf} for 1992 BF. In cases where the spin state is well characterized, usually through the combination of radar imaging and photometric light curves, the bulk density of the object is correlated with the thermal properties and mutual constraints can be inferred, as was the case for Golevka \citep{chesley_golevka}. \citet{rozitis_2013_apollo} were able to jointly model the measured Yarkovsky and YORP effects on (1862) Apollo, and thereby constrain a number of the body's physical characteristics, including axis ratios, size, albedo, thermal inertia and bulk density. In the case of Bennu, the thermal inertia is known from infrared observations \citep{emery_bennu_2014, muller_2012_rq36}, and so we are able to directly estimate the mass and bulk density.


\section{Observational Data and Treatment}\label{sec:data}

\subsection{Optical Astrometry}

We use the 569 RA-DEC astrometric measurements available from the Minor Planet Center from 1999-Sep-11.4 to 2013-Jan-20.1. We apply the star catalog debiasing algorithm introduced by \citet{cbm_2010}, and data weights are generally based on the astrometric weighting scheme proposed in Sec. 6.1 of that paper. In some cases there is an excess of observations from a single observatory in a single night. In such cases we relax the weights by a factor of about $\sqrt{N/5}$, where $N$ is the number of observations in the night. This reduces the effect of the particular data set to a level more consistent with the typical and preferred contribution of 3--5 observations per night.

\begin{figure}[t]
\begin{center}
\includegraphics[width=5in]{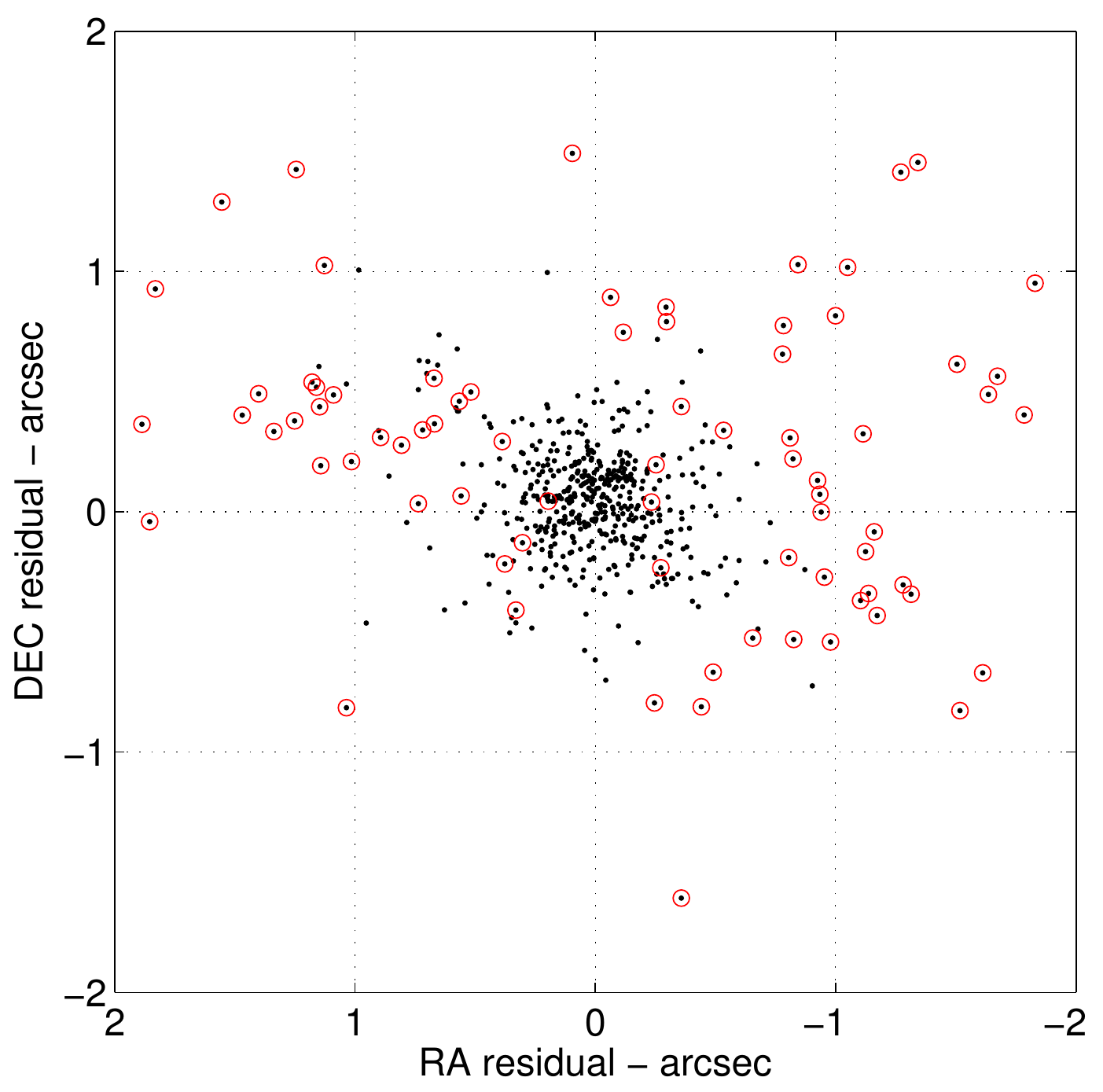}
\caption{Depiction of Bennu postfit residuals for JPL solution
  87. Deleted observations are depicted with circles around the
  points. In addition to the observations shown here, there are 14
  deleted observations outside the plot boundaries.}
\label{fig:resids}
\end{center}
\end{figure}

Considerable care was taken in identifying outlier observations to be deleted as discordant with the bulk of the observations. From among the 569 available observations from 43 stations, we reject 91 as outliers, leaving 478 positions from 34 stations in the fits. Figure~\ref{fig:resids} depicts the postfit plane of sky residuals, highlighting the deleted data. There are an additional 14 observations, all deleted, that are not depicted in Fig.~\ref{fig:resids}  because they fall beyond the limits of the plot. The manual rejection approach often deletes an entire batch of data if it appears biased in the mean, thus some of the deleted points in Fig.~\ref{fig:resids} do not show significant residuals. On the other hand, some observations are de-weighted relative to the others, and in some cases these are not deleted, despite the raw residuals being larger than some rejected observations. In Sec.~\ref{sec:outlier} we discuss the dependency of the ephemeris prediction on the outlier rejection approach.

\subsection{Radar Astrometry}

The time delay and Doppler shift of radar echoes from Bennu were
measured in 1999, 2005 and 2011. Radar astrometry was obtained at both Arecibo and Goldstone as detailed in Table~\ref{tab:radar}. The delay observations in the table correspond
to the round-trip light travel time from the nominal telescope position
to the center of mass of the object, and thus they are often referred
to as range measurements. Doppler measurements in the table reflect
the frequency shift between the transmit and receive signals due to the line-of-sight velocity of the object. The use
of radar delay and Doppler measurements in asteroid orbit
determination was introduced by \citet{yeomans_radar_1992}.

Delay uncertainties arise from the finite resolution of the imaging of
0.05-0.125 $\mu$s/px \citep{nolan_etal_2013}, uncertainty in the shape
modeling (to determine the center of mass from the observed echo
power) of 10-20 m, equivalent to about 0.1 $\mu$s, and systematic
calibration, including uncertainties in the position of the telescope
and light travel within the telescope optics. Because we have a shape
model of Bennu that directly relates the individual range
observations to the center of figure of the model
\citep{nolan_etal_2013}, the systematic uncertainties dominate the
range uncertainty in the 1999 and 2005 observations, and are assigned
conservative values of 1.0 and 0.5~$\mu$s (respectively). In 2011,
Bennu was much farther away than the previous observations,
and the uncertainty of 2~$\mu$s is from the pixel scale of the
observations. Doppler uncertainties are taken to be 1~Hz at 2380~MHz,
about 1/4 of the total rotational Doppler width of the object, and are
based on the uncertainty of estimating the position of the center of
mass of the spectra.

\begin{figure}[t]
\begin{center}
\includegraphics[width=5in]{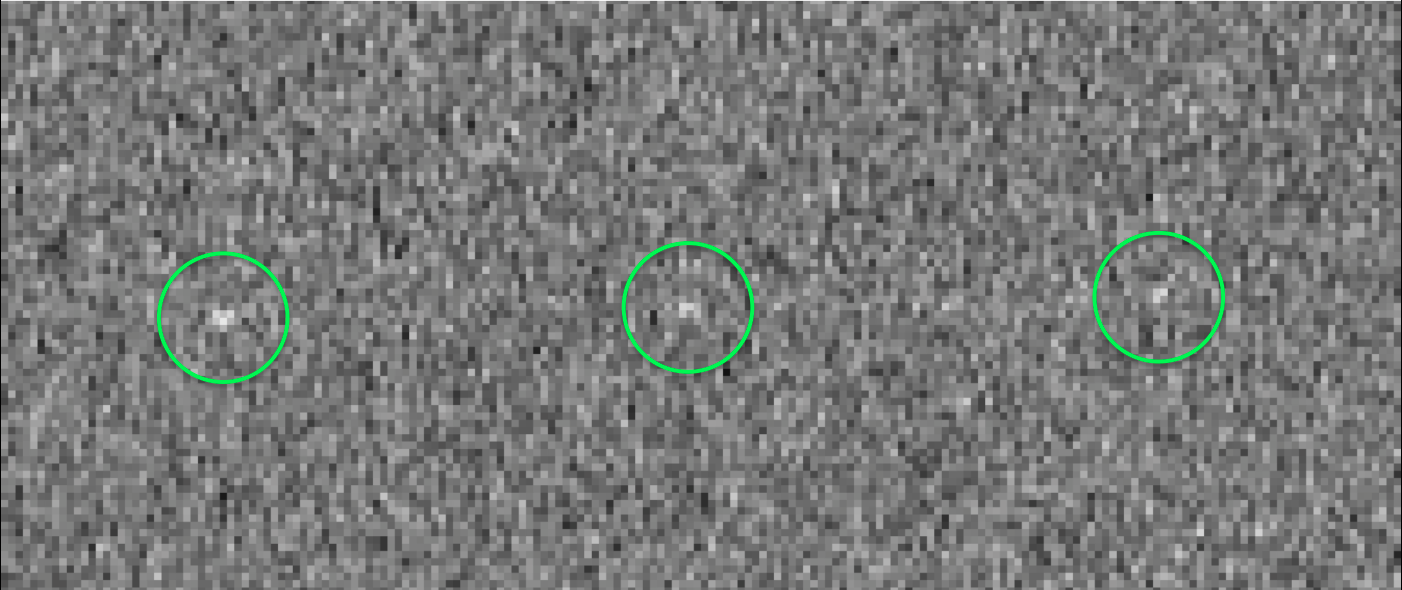}
\caption{Arecibo delay-Doppler detections of Bennu from 2011-Sep-27,28,29. Doppler frequency increases to the right and delay increases upwards. Image resolution is $1\, \mathrm{Hz} \times 1\,\mu\mathrm{s}$.}
\label{fig:arecibo}
\end{center}
\end{figure}

The 2011 observations (Fig.~\ref{fig:arecibo} and Table~\ref{tab:2011_radar}) were of too low resolution and SNR to be useful for shape modeling and were obtained solely for improving our knowledge of the orbit of Bennu. The 2-$\mu$s (300~m) resolution was chosen to be the finest resolution that would maximize the SNR of the observations by including all of the echo power from the 250-m radius asteroid in one or two range bins. The asteroid was visible with a $\rm{SNR} > 3$  on each of the three observing dates at consistent delay and Doppler offsets from the a priori ephemeris used in the data taking.

The 2011 radar observations of Bennu, which enabled the results of
the present paper, almost never happened. The two-million-Watt,
65,000-Volt ``power brick'' that supplies the electricity for the
Arecibo Planetary Radar system failed in late 2010, and was finally
repaired on September 15, 2011. Because of the critical schedule for
Bennu observations, in a space of seven days the 16-ton unit was
trucked 800 miles from Pennsylvania to Florida, shipped to Puerto
Rico, trucked again, and lifted into place with a crane. The system
was reconnected and recommissioned in four days, after nearly a year
of down-time, just in time to perform the observations on the last
possible dates of September 27--29, and just as the prime contractor
managing the Arecibo Observatory was changing (on October 1), after 45
years of operation by Cornell University, so that most observatory
operations were frozen for the transition.

\begin{table}[ht]
\caption{Radar Astrometry for (101955) Bennu.\label{tab:radar}}
\small
\begin{center}
\begin{tabular}{lrcc}
\hline
Date \& Time & Measurement     &  Uncertainty      & Station \\
(UTC - Receive) & ($\mu$sec or Hz)     &  ($\mu$sec or Hz)      & \\
\hline\\
Delay		    &		    & 	    &	        \\
2011-09-29 11:55:00 & 202378520.04  &  2.0  & Arecibo   \\
2011-09-28 11:08:00 & 199711477.27  &  2.0  & Arecibo   \\
2011-09-27 11:39:00 & 197293588.79  &  2.0  & Arecibo   \\[0.2cm]
2005-10-02 14:10:00 &  57762582.67  &  0.5  & Arecibo   \\
2005-10-02 12:55:00 &  57594268.2   &  0.5  & Arecibo   \\
2005-09-28 13:35:00 &  45734943.4   &  0.5  & Arecibo   \\
2005-09-28 11:57:00 &  45550976.4   &  0.5  & Arecibo   \\
2005-09-20 11:24:00 &  33024222.68  &  0.5  & Arecibo   \\
2005-09-20 09:09:00 &  33024251.3   &  0.5  & Arecibo   \\
2005-09-19 12:50:00 &  33215505.86  &  1.0  & Goldstone \\
2005-09-18 12:20:00 &  33873463.    &  1.0  & Goldstone \\
2005-09-16 09:27:00 &  36609699.09  &  0.5  & Arecibo   \\
2005-09-16 08:45:00 &  36658796.03  &  0.5  & Arecibo   \\[0.2cm]
1999-10-01 13:40:00 &  35441297.    &  1.0  & Goldstone \\
1999-09-25 12:55:00 &  17785960.83  &  1.0  & Arecibo   \\
1999-09-25 11:09:00 &  17634668.28  &  1.0  & Arecibo   \\
1999-09-24 12:23:00 &  15955075.55  &  1.0  & Arecibo   \\
1999-09-24 10:26:00 &  15838961.63  &  1.0  & Arecibo   \\
1999-09-23 11:28:00 &  14846130.16  &  1.0  & Arecibo   \\
1999-09-23 09:36:00 &  14800106.19  &  1.0  & Arecibo   \\
1999-09-23 09:30:00 &  14820631.    &  5.0  & Goldstone \\
1999-09-21 10:20:00 &  15418454.    & 10.0  & Goldstone \\
		    &		    & 	    &	        \\
Doppler		    &		    & 	    &	        \\
2011-09-29 11:55:00 & -72841.0156   &  1.0  & Arecibo   \\
2011-09-28 11:08:00 & -68554.7858   &  1.0  & Arecibo   \\
2011-09-27 11:39:00 & -66400.2088   &  1.0  & Arecibo   \\[0.2cm]
2005-09-28 12:00:00 & -73137.0697   &  1.0  & Arecibo   \\
2005-09-20 09:06:00 &   2631.7168   &  1.0  & Arecibo   \\
2005-09-16 08:44:00 &  47170.7359   &  1.0  & Arecibo   \\[0.2cm]
1999-09-21 09:00:00 & 135959.        & 5.0  & Goldstone \\
\hline
\end{tabular}
\end{center}
{\em Notes:} \begin{itemize}
\item Transmit frequency is 2.38 GHz at Arecibo and 8.56 GHz
at Goldstone.
\item All measurements are referenced to the body center of mass.
\item Measurements are also available online at {\tt http://ssd.jpl.nasa.gov/?radar}.
\end{itemize}
\end{table}

\begin{table}[ht]
\caption{Radar observations of Bennu from 2011.  Each line gives the UTC date and start/stop times, the number of transmit-receive cycles (runs), and the direction and distance to Bennu at the mid-epoch on each date. All observations used a 2 microsecond baud (corresponding to a range resolution of 300 meters), an 8191-length code, and JPL/Horizons orbital solution 70.\label{tab:2011_radar}}
\small
\begin{center}
\begin{tabular}{ccccccc}
\hline
UTC Date     & Start Time & Stop Time & Runs & RA ($^\circ$) & DEC ($^\circ$) & Distance (au)\\
\hline
2011-Sep-27  & 10:30:36 & 12:53:45  &  21  & 114.3 & +28.3  &  0.1977\\
2011-Sep-28  & 10:34:40 & 12:56:31  &  10  & 116.2 & +28.5  &  0.2002\\
2011-Sep-29  & 10:38:31 & 13:02:29  &  18  & 118.1 & +28.6  &  0.2028\\
\hline
\end{tabular}
\end{center}
\end{table}

\section{Orbit Determination and Dynamical Model}\label{sec:od}

We have updated the orbit determination for Bennu based on the
observational data set described above. These
orbital position measurements place extraordinary constraints on the
orbit determination, and thus we must pay careful attention to the
fidelity of force models, observation models and numerical
integration. Our dynamical model includes direct solar radiation
pressure and the thermal re-emission of absorbed solar radiation
(i.e., the Yarkovsky effect). Besides the gravitational acceleration
of the Sun, we include Newtonian perturbations by the eight planets, the Moon, Pluto and 25 selected main belt asteroids. We consider the
oblateness term of the Earth's geopotential and full relativistic
perturbations from the Sun, eight planets and the Moon.

As shown by \cite{giorgini_1950da}, who studied the potential impact
of 29075 (1950 DA) in the year 2880, other potential dynamical
perturbations, such as galactic tide, solar mass loss and solar
oblateness, are too slight to affect our results. This is because
these small effects, which were not important for 1950 DA, will be even
less significant for Bennu due to the much shorter time
interval. 

\subsection{Yarkovsky Effect}\label{sec:yarko}

The Yarkovsky effect is a key consideration when fitting an orbit for Bennu \citep{milani_etal_rq36}. This slight nongravitational acceleration arises from the anisotropic re-emission at thermal wavelengths of absorbed solar radiation \citep{bottke_etal06}. The component of the thermal recoil acceleration in the transverse direction acts to steadily increase or decrease the orbital energy, leading to a drift in semimajor axis $da/dt$ that accumulates quadratically with time in the orbital longitude of the asteroid. For a uniform, spherical asteroid on a known orbit, the drift rate depends on the physical characteristics of the asteroid according to
$$
\frac{da}{dt} \propto \frac{\cos \gamma}{\rho D},
$$ 
where $\gamma$ is the obliquity of the asteroid equator with respect to its orbital plane, $\rho$ is the bulk density of the asteroid, and $D$ is the effective diameter. Additionally, $da/dt$ depends in a nonlinear and often nonintuitive way on the asteroid rotation period $P$ and the surface material properties, namely thermal inertia $\Gamma$, infrared emissivity $\epsilon$ and Bond albedo $A$  \citep{vok_etal_2000}.

We have three models available to us for computing thermal accelerations on Bennu. The first, and most straightforward, is to simply apply a transverse acceleration of the form $A_T\times (r/1\,{\rm au})^{-d}$, where $A_T$ is an estimable parameter, $r$ is the heliocentric distance and the exponent is typically assumed as $d=2$ to match the level of absorbed solar radiation. Given an estimated value of $A_T$ and the assumed value of $d$, one can readily derive the time-averaged $da/dt$ using Gauss' planetary equations \citep{farnocchia_yarkolist_2013}. This approach, which we term the {\em transverse model}, is computationally fast and captures the salient aspects of the thermal recoil acceleration. Importantly, it requires no information about the physical characteristics or spin state of the asteroid, and so it can be implemented readily in cases where only astrometric information is available \citep[e.g.,][]{1992bf, chesley_acm08, nugent_etal_2012,farnocchia_yarkolist_2013}.

For Bennu we find numerically that the exponent $d=2.25$ provides the best match to the transverse thermal acceleration derived from the thermal re-emission models described below. This result can also be computed analytically using a simplified model with the technique described in Appendix A. Using the transverse model with $d=2.25$ we derive JPL solution 87 (Table~\ref{tab:nominal}), which serves as a reference solution as we investigate the effect of various model variations on the orbit. 

JPL solution 87 yields a Yarkovsky drift estimate that compares well with the corresponding result from \citet{milani_etal_rq36}, who used observations only through mid-2006 and found $da/dt = (-15 \pm 9.5)\times 10^{-4}$ au/Myr, which was judged to be a weak detection of the nongravitational acceleration. Using the same fit span (1999--2006) from the current data set we now find $da/dt = (-22.9 \pm 5.3)\times 10^{-4}$ au/Myr. The change in the estimate relative to that of \citet{milani_etal_rq36} is due in large part to the use of star catalog debiasing \citep{cbm_2010}, while the improved precision is due to the higher accuracy and quantity of radar delay measurements obtained through re-measurement of the 1999 and 2005 Arecibo observations, as well through as the use of tighter weights on the optical data proposed by \citet{cbm_2010}. Incorporating the subsequent optical observations through 2013 leads to $da/dt = (-21.3\pm 4.6)\times 10^{-4}$ au/Myr. Finally, adding the 2011 Arecibo radar astrometry reduces the uncertainty by nearly a factor 50, leading to the current best estimate $da/dt = (-19.0 \pm 0.1)\times 10^{-4}$ au/Myr. We note that the new formal uncertainty on $da/dt$ is 0.5\%, by far the most precise Yarkovsky estimate available to date. As well, the uncertainty on the semimajor axis $a$ is 6 m, the lowest value currently found in the asteroid catalog. This low uncertainty is primarily a reflection of the current precision of the orbital period (2 ms) rather than an indication of the uncertainty in the predicted asteroid position, which is at the level of a few kilometers during the fit span.

\begin{table}[t]
\caption{JPL orbit Solns. 85 and 87 for Bennu, Ecliptic J2000 Frame. \label{tab:nominal}}
\small
\begin{center}
\begin{tabular}{ll}
\hline
\multicolumn{2}{l}{{\bf Solution 85} (Nonlinear Yarkovsky Model)}\\

Epoch                         & 2011-Jan-1.0 TDB              \\
Semimajor axis ($a$)          & $1.126391025996(42)$ au       \\
Eccentricity ($e$)            & $0.203745112(21)$             \\
Perihelion dist. ($q$)        & $0.896894360(24)$ au          \\
Perihelion time ($t_p$)       & 2010-Aug-$30.6419463(30)$ TDB \\
Long. of Asc. Node ($\Omega$) &  $2.0608668(37)^\circ$         \\
Arg. of perihelion ($\omega$) & $66.2230699(55)^\circ$        \\
Inclination ($i$)             &  $6.0349391(27)^\circ$         \\
Bulk Density ($\rho$)         & $1181.1(6.3)$ kg/m$^3 \dagger$ \\
$\chi^2\;\ddag$               & 68.37                         \\[0.5cm]

\multicolumn{2}{l}{{\bf Solution 87} (Transverse Yarkovsky Model, $d=2.25$)}\\
Epoch                         & 2011-Jan-1.0 TDB              \\
Semimajor axis ($a$)          & $1.126391026404(40)$ au       \\
Eccentricity ($e$)            & $0.203745114(21)$             \\
Perihelion dist. ($q$)        & $0.896894358(24)$ au          \\
Perihelion time ($t_p$)       & 2010-Aug-$30.6419468(30)$ TDB \\
Long. of Asc. Node ($\Omega$) & $2.0608670(37)^\circ$         \\
Arg. of perihelion ($\omega$) & $66.2230705(55)^\circ$        \\
Inclination ($i$)             & $6.0349388(27)^\circ$         \\
Transverse accel. ($A_T$)     & $-4.618(24) \times 10^{-14}$ au/d$^2$ \\
$\chi^2\;\ddag$               & 68.73                         \\

\hline
\end{tabular}
\end{center}
{\em Notes}: 

Numbers in parentheses indicate the $1\sigma$ formal uncertainties of the corresponding (last two) digits in the parameter value.

$\dagger$ The bulk density uncertainty is marginal only with respect to the orbital elements, and is conditional with respect to the physical parameters that can affect the thermal modeling. In particular, the uncertainties in effective diameter, thermal inertia and obliquity are not captured here, and these lead to a marginal uncertainty an order of magnitude greater. See Fig.~\ref{fig:rho} and the discussion in Sec.~\ref{sec:rho}.

$\ddag$ $\chi^2$ denotes the sum of squares of normalized postfit residuals.

\end{table}

Both our second (linear) and third (nonlinear) Yarkovsky acceleration models employ heat transfer models of different levels of fidelity in order to predict the surface temperature and associated re-emission of thermal energy. The {\em linear model} utilizes linearized heat transfer equations on a rotating homogeneous sphere, closely following the development given by \citet{vok_etal_2000} for both the diurnal and seasonal components of the Yarkovsky effect. The linear model requires knowledge of the spin orientation and rate, asteroid diameter and thermal inertia, but does not allow for shape effects such as self-shadowing and self-heating, which are generally considered minor. The linear model assumes a sphere, and so oblateness effects are not captured. This is relevant because the cross-sectional area receiving solar radiation is increased for an equal volume sphere relative to that of an oblate body, and thus the force derived with the linear model is enhanced relative to the nonlinear model. This in turn leads to an increased estimate of the bulk density as we shall see later.

The {\em nonlinear model} is the highest fidelity Yarkovsky force
model that we apply to the orbit determination problem. This approach
solves the nonlinear heat transfer equation on a finite-element mesh
of plates or facets that models the \citet{nolan_etal_2013} asteroid shape. The
approach is described in more detail by
\citet{capek_belgrade}, but we summarize it here. For each facet on
the asteroid shape model, the nonlinear heat transfer problem is
solved while the asteroid rotates with a constant spin rate and
orientation and revolves along a frozen, two-body heliocentric
orbit. A uniform temperature distribution is assumed at start-up and
the temperature and energy balance between absorbed, conducted and
re-radiated radiation for each facet is solved as a function of time. The heat transfer problem is treated as one-dimensional, i.e., the temperature for a given facet depends only on the depth below the facet. There is no conduction across or between facets. After several
orbital revolutions the temperature profile from
revolution to revolution converges for each plate. Following
convergence, diurnal averaging of the vector sum of the thermal emission over the body
yields the force of thermal emission as a function of orbital
anomaly. Given the shape model volume and an assumed bulk density, the
mass can be computed and from this the thermal recoil
acceleration. This ultimately leads to a lookup table of acceleration
as a function of true anomaly that is interpolated during the
high-fidelity orbital propagation.

\begin{figure}[t]
\begin{center}
\includegraphics[width=5in]{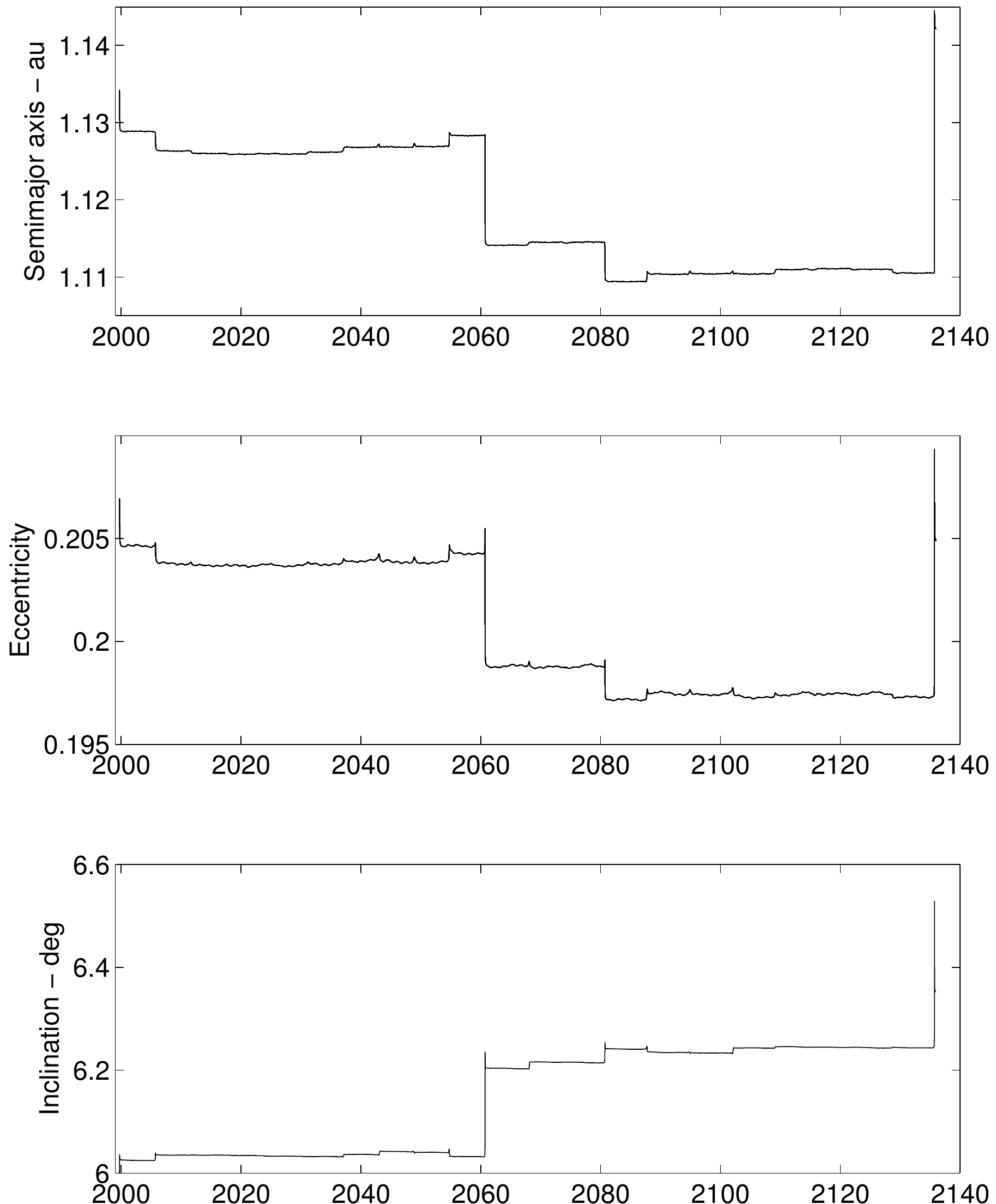}
\caption{Time history of the osculating Bennu orbital elements, $a$, $e$ and
$i$, through 2136. The effects of Earth encounters are evident and can
be cross-referenced with Table~\ref{tab:det_ca}.}
\label{fig:elem_hist}
\end{center}
\end{figure}

\begin{figure}[t]
\begin{center}
\includegraphics[width=5in]{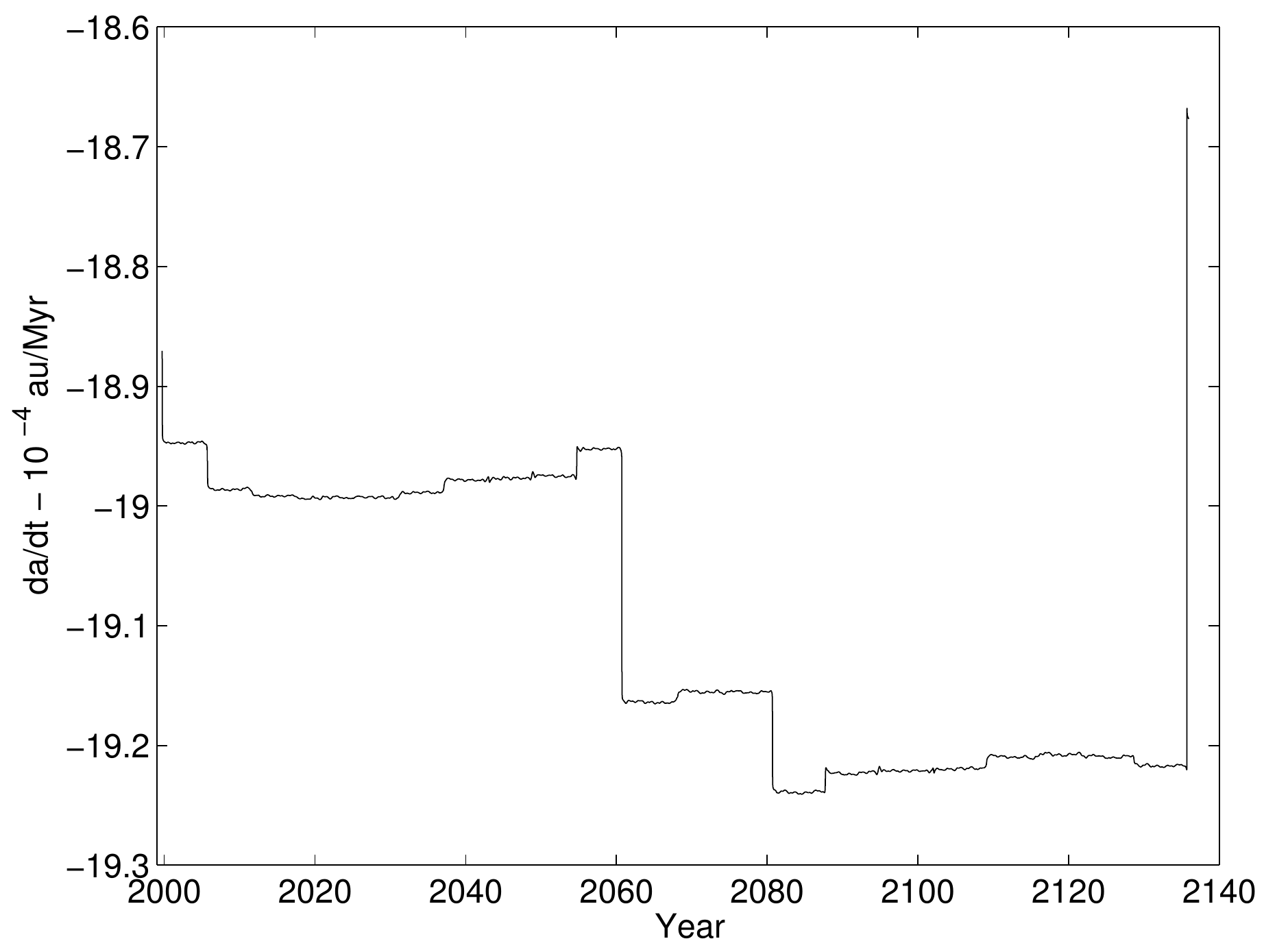}
\caption{Time history of the Bennu semimajor axis drift $da/dt$
through 2136. The variations arise from orbital changes induced by
Earth encounters as depicted in Fig.~\ref{fig:elem_hist}.}
\label{fig:dadt_hist}
\end{center}
\end{figure}

The nonlinear model was previously used with asteroid (6489) Golevka
\citep{chesley_golevka}, but at that time the acceleration table was
for a frozen orbit, which turns out to be an unacceptable
approximation for Bennu. Figure~\ref{fig:elem_hist} shows the orbital element
variations into the future due to planetary perturbations and Fig.~\ref{fig:dadt_hist} reveals the associated
variation in the average $da/dt$, which is clearly significant
relative to the $0.10\times 10^{-4}$ au/Myr uncertainty. As a result
of this analysis we have implemented an enhancement to the nonlinear
model that corrects the tabulated accelerations for variations in
orbital elements. The approach is to compute the Yarkovsky force
vector from a linearized expansion about a central, reference orbit according to
$$
\vec F_Y(a, e; f_i) = \vec F_Y(a_0, e_0; f_i) + 
    \frac{\partial \vec F_Y(a_0, e_0; f_i)}{\partial a} (a - a_0) + 
    \frac{\partial \vec F_Y(a_0, e_0; f_i)}{\partial e} (e - e_0). 
$$ 

Here $\vec F_Y$ is the thermal acceleration in the orbit plane frame so that  variations of the Keplerian Euler angle orbital elements ($\omega$, $\Omega$, $i$) do not affect the computation; $\vec F_Y$ is rotated to the inertial frame during the propagation. The $f_i$ are the true anomaly values in the tabulation, $a_0$ and $e_0$ are the values for the reference orbit. The partial derivatives are also tabulated after they are derived by finite differences based on a series of pre-computed lookup tables for varying orbits $\vec F_Y(a_0 \pm\delta a, e_0 \pm\delta e; f_i)$.

When computing an orbit with the linear or nonlinear model we use the
physical parameters listed in Table~\ref{tab:rq36_props}, and for the
nonlinear model we also use the \citet{nolan_etal_2013} shape
model. The bulk density $\rho$ is estimated as a free
parameter. Because the semimajor axis drift $da/dt$ is constrained by
the observations at the 0.5\% level, any variations in the force
computed by the thermal models manifests as a variation in the
estimated bulk density. This is discussed in greater detail in
Sec.~\ref{sec:rho}. 

\begin{table}[t]
\caption{Physical characteristics of (101955) Bennu and associated marginal uncertainty in estimate of bulk density $\rho$. Tabulated error bars represent assumed $1\sigma$ uncertainties. \label{tab:rq36_props}}
\small
\begin{center}
\begin{tabular}{llll}
\hline
 Parameter             & Value \& Uncertainty                                &Ref.& $\rho$ Uncert.  \\
\hline
 Thermal inertia       & $\Gamma=310\pm70\;\rm{J\,m^{-1}\,s^{-0.5}\,K^{-1}}$ & A  & $+2.1\%/{-4.1}\%$     \\
 Cross-sectional Area/Volume & $A/V= 3.06\pm 0.06\times 10^{-3}\,\rm{m}^{-1}$     & B  & $\pm 2.0\%$       \\
 Obliquity of equator  & $\gamma= 175^\circ \pm 4^\circ$                             & B  & $+0.4\%/{-0.9}\%$ \\
 Surface emissivity    & $\epsilon= 0.90\pm 0.05$                            & A  & $\pm 0.3\%$     \\
 Bond albedo           & $A = 0.017 \pm 0.002$                               & A  & $\pm 0.2\%$     \\
 Rotation period       & $4.29746 \pm 0.002$ h                               & B  & $\pm 0.0\%$       \\
\hline
\end{tabular}
\end{center}
{\em References}: A--- \citet{emery_bennu_2014}, B--- \citet{nolan_etal_2013}
\end{table}

\begin{figure}[t]
\begin{center}
\includegraphics[width=5in]{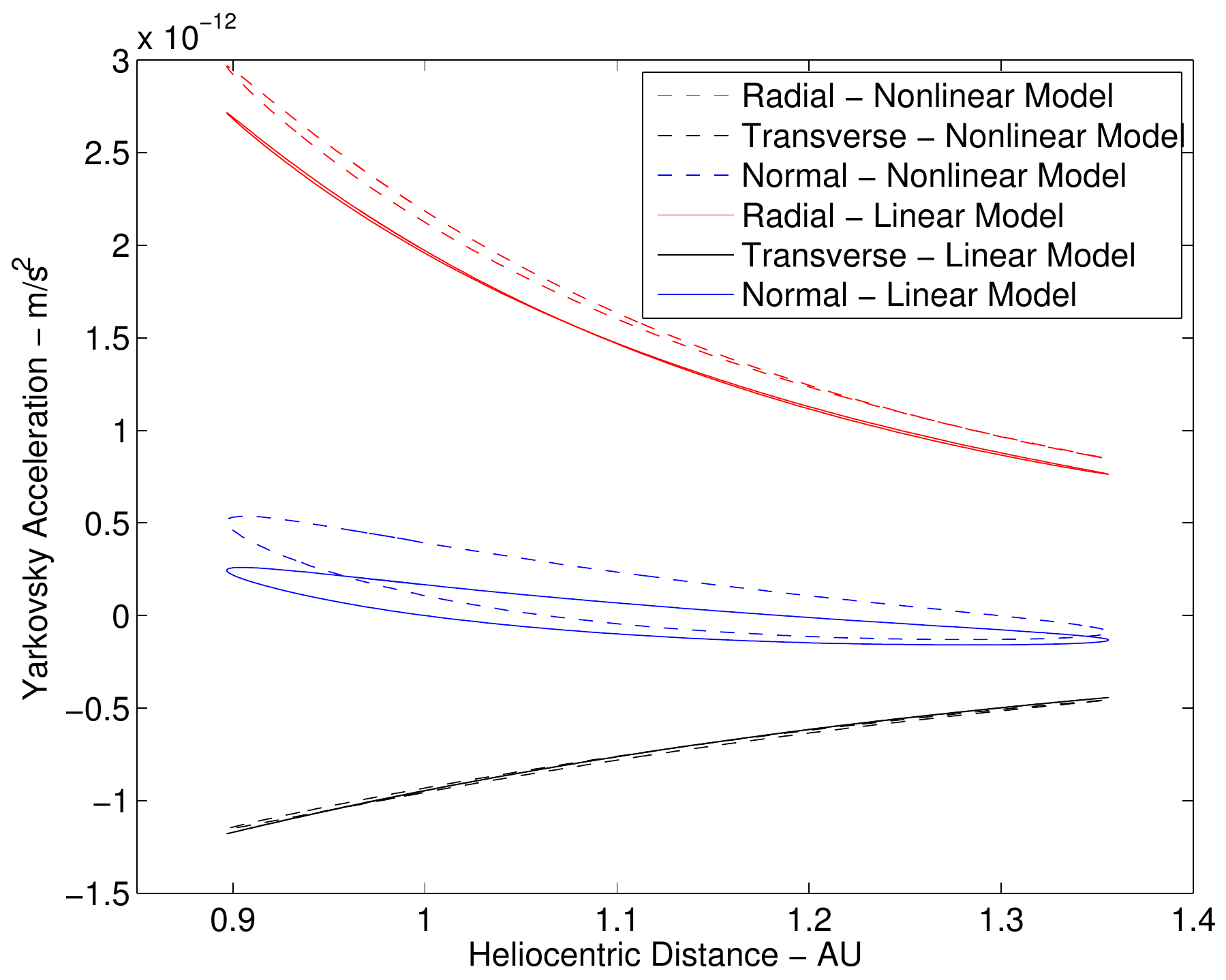}
\caption{Yarkovsky accelerations as a function of heliocentric
distance, according to the linear and nonlinear Yarkovsky models.}
\label{fig:joint_yarko}
\end{center}
\end{figure}

\begin{figure}[t]
\begin{center}
\includegraphics[width=5in]{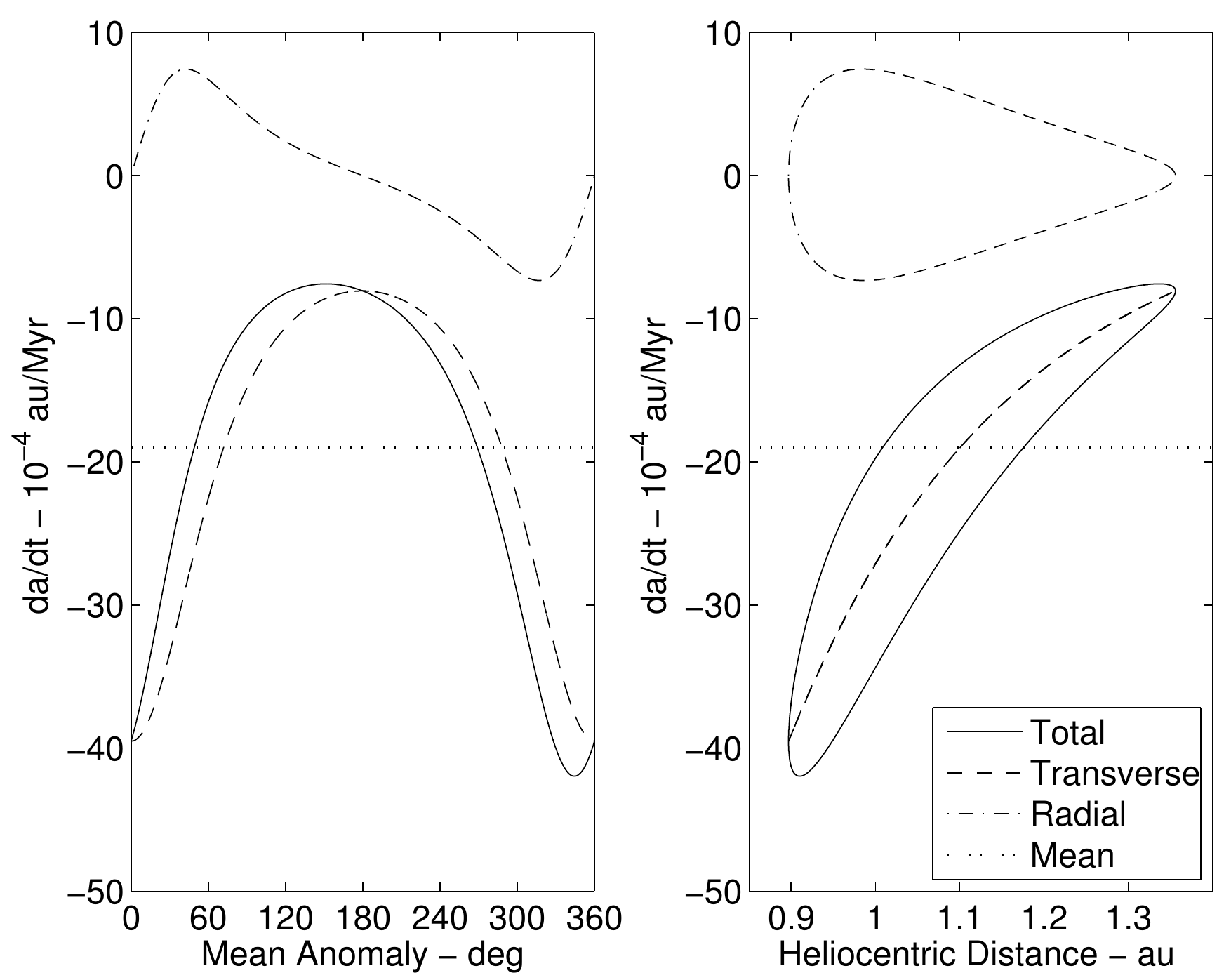}
\caption{The history of $da/dt$ stemming from the transverse and
radial components of the Yarkovsky acceleration. The linear Yarkovsky model is depicted.}
\label{fig:dadt}
\end{center}
\end{figure}

JPL solution 85 (Table~\ref{tab:nominal}) uses the nonlinear model, and is assumed to be our most accurate orbital solution. To indicate the differences between the two models, Fig.~\ref{fig:joint_yarko} depicts the estimated thermal recoil accelerations from the nonlinear and linear models. The plot reveals an excellent agreement between the two models in the transverse acceleration, which is to be expected since the transverse component is constrained by the observed orbital runoff and the associated semimajor axis drift. The radial and normal (out-of-plane) accelerations show a good but imperfect agreement, with linear model accelerations being noticeably reduced relative to the nonlinear model. 

Figure~\ref{fig:dadt} shows how the radial and transverse accelerations in the linear model affect the instantaneous and average values of $da/dt$, as derived from the classical Gauss planetary equations. During a given orbit, the variations in the instantaneous drift rate are much larger from the transverse component. The normal component of acceleration does not affect the semimajor axis. While the radial component of acceleration does lead to variations in semimajor axis during an orbital period, in the mean the radial term does not contribute to semimajor axis drift, which is a classical result if the radial accelerations are symmetric about perihelion. However, Fig.~\ref{fig:joint_yarko} reveals that symmetry is not necessarily present in this case. There is a slight hysteresis in the radial acceleration profile for the linear model, but because the curve crosses itself the integrated area under the curve in one orbit nets to approximately zero. In contrast, the nonlinear radial acceleration has a more significant hysteresis that does not sum to zero, and thus the radial component of acceleration actually contributes to $da/dt$ in the mean. This behavior is presumably associated with the fact that thermal energy penetrates more deeply below the asteroid surface around perihelion when the absorbed radiation is greatest, which leads to greater thermal emission post-perihelion that pre-perihelion. We find that in the nonlinear model the radial acceleration increases $da/dt$ by 0.3\%, which is not negligible relative to the 0.5\% precision of the estimate. The result is that the transverse component must contribute 0.3\% more in magnitude to compensate. With the linear model the radial contribution to $da/dt$ is 60 times less.

Table~\ref{tab:models} shows the variation in estimated $da/dt$ associated with the different Yarkovsky models. JPL solution 87 ($d=2.25$) is the reference solution for the comparison, and the linear and nonlinear models yield $da/dt$ values within 0.03\%, less than a tenth of the formal uncertainty. The result for the typical default value $d=2$ is also tabulated and agrees well. This is not surprising since the astrometry provides a strong constraint on $da/dt$ that the models must accommodate. At this level of precision, the averaged $da/dt$ may not be the best means of quantifying the Yarkovsky effect because the mean value changes as the orbit undergoes strong planetary perturbations (Figs.~\ref{fig:elem_hist} and \ref{fig:dadt_hist}). Nonetheless, it is informative when comparing objects and assessing the scale of the Yarkovsky effect and so we continue to use it here.

As discussed in Sec.~\ref{sec:future}, Bennu will have a close approach to Earth in 2135 at around the lunar distance. Table~\ref{tab:models} also lists the variation in the 2135 $b$-plane coordinates $(\xi_{2135}, \zeta_{2135})$ associated with the different Yarkovsky models. We describe these coordinates more fully later, but the salient point is that $\Delta\zeta_{2135}$ reveals the importance of the model variation for long term predictions, while $\Delta da/dt$ reflects the relevance to the orbital estimate over the fit span from 1999-2013. 


In addition to the Yarkovsky effect, our dynamical model also includes another nongravitational perturbation related to solar radiation, namely direct solar radiation pressure \citep[SRP,][]{vok_mil_2000}. Based on the Nolan shape model and the mass estimate discussed in Sec.~\ref{sec:rho}, we assume an area-to-mass ratio of $2.59\times 10^{-6}\, \rm{m^2/kg}$, which leads to an acceleration of $1.2\times10^{-11}\, \rm{m/s^2}$ at 1 au, an order of magnitude greater than the radial acceleration from thermal re-emissions (see Fig.~\ref{fig:joint_yarko}). Reflected radiation pressure is negligible due to the 1.7\% Bond albedo of the body \citep{emery_bennu_2014}. Even though the acceleration of SRP is several times greater than that from thermal re-emission, it has little effect on the orbital predictions as it is perfectly aliased with the solar gravity. Turning SRP on and off changes the estimated semimajor axis (by $\sim 67\sigma$) but leaves the mean motion unchanged. Thus there is only a minor effect on the trajectory from eliminating solar radiation pressure from the force model (by assuming an area-to-mass ratio of zero), as can be seen in Table~\ref{tab:models} under the entry labeled Area/Mass$=0$.

\begin{table}[t]
\caption{Dynamical Effect of Several Model Variations. The columns indicate the type of model variation and the associated change in semimajor axis drift rate $da/dt$ or 2135 $b$-plane coordinates $(\xi_{2135}, \zeta_{2135})$. Tabulated $\Delta$ values are with respect to JPL solution 87 (Table~\ref{tab:nominal}), for which $da/dt = -18.973 \times 10^{-4}$ au/My at epoch 2011-Jan-1.0 and $(\xi_{2135}, \zeta_{2135}) = (-125\,932\,\mathrm{km}, 281\,597\,\mathrm{km})$. \label{tab:models}}
\small
\begin{center}
\begin{tabular}{l.rrl}
\hline
 Model   & \multicolumn{1}{c}{$\Delta da/dt$} & $\Delta\xi_{2135}$ & $\Delta\zeta_{2135}$ & Remarks \\
                  & \multicolumn{1}{c}{($10^{-4}$ au/My)} & (km) & (km) &  \\
\hline
\multicolumn{5}{l}{\em Yarkovsky Model}\\
 \hspace{3mm}Nonlinear         &  -0.004 &    -43&   6251  & Soln. 85  \\
 \hspace{3mm}Linear            &   0.006 &    -16&   2096  & Soln. 86  \\
 \hspace{3mm}$d=2.25$          &   0.000 &      0&      0  & Soln. 87  \\
 \hspace{3mm}$d=2.00$          &   0.006 &    -21&   3423  &  \\[2mm]
\multicolumn{5}{l}{\em Asteroid Perturbations}\\
 \hspace{3mm}25 Perturbers     &  0.000 &      0&      0  & Soln. 87  \\
 \hspace{3mm}BIG-16 only       & -0.004 &      2&   -409  &  \\
 \hspace{3mm}CPVH only         & -0.010 &     22&  -3714  &  \\[2mm]
\multicolumn{5}{l}{\em Earth Oblateness Limit}\\
 \hspace{3mm}10.     au        &   0.000 &      0&    -70 &  \\
 \hspace{3mm}1.      au        &   0.000 &      0&      0 & Soln. 87  \\
 \hspace{3mm}0.1     au        &   0.000 &      0&    -25 &  \\
 \hspace{3mm}0.01    au        &  -0.004 &     11&  -1822 &  \\
 \hspace{3mm}0.001   au        &  -0.004 &     10&  -1703 &  \\[2mm]
\multicolumn{5}{l}{\em Relativity Model}\\
 \hspace{3mm}Full EIH          &  0.000 &      0&      0 & Soln. 87  \\
 \hspace{3mm}Basic Sun Model   &  0.305 &  -1128& 168469 &  \\
 \hspace{3mm}EIH Sun only      &  0.295 &  -1069& 160086 &  \\
 \hspace{3mm}w/o Mercury       & -0.001 &      1&   -240 &  \\
 \hspace{3mm}w/o Venus         &  0.017 &    -54&   8954 &  \\
 \hspace{3mm}w/o Earth         &  0.291 &  -1059& 159130 &  \\
 \hspace{3mm}w/o Mars          &  0.000 &      0&    -43 &  \\
 \hspace{3mm}w/o Jupiter       & -0.012 &     46&  -7652 &  \\
 \hspace{3mm}w/o Saturn        & -0.004 &     11&  -1859 &  \\
 \hspace{3mm}w/o Uranus        &  0.000 &     -1&    168 &  \\
 \hspace{3mm}w/o Neptune       &  0.000 &     -0&     60 &  \\
 \hspace{3mm}w/o Moon          &  0.004 &    -11&   1801 &  \\[2mm]
\multicolumn{5}{l}{\em Outlier Rejection}\\
 \hspace{3mm}$\chi_{\rm{rej}}=3  $&  0.049 &   -537&  84670  &  7 del.  \\
 \hspace{3mm}$\chi_{\rm{rej}}=2  $&  0.026 &   -414&  65916  & 15 del.  \\
 \hspace{3mm}$\chi_{\rm{rej}}=1.5$&  0.018 &   -243&  39502  & 24 del.  \\
 \hspace{3mm}$\chi_{\rm{rej}}=1  $&  0.003 &    -13&   2220  & 49 del.  \\
 \hspace{3mm}Manual              &  0.000 &      0&      0  & 91 del., Soln. 87 \\[2mm]
\multicolumn{5}{l}{\em Integration Tolerance}\\
 \hspace{3mm}$10^{-16}$        &   0.000 &     -0&     60  &  \\
 \hspace{3mm}$10^{-15}$        &   0.000 &      0&      0  & Soln. 87  \\
 \hspace{3mm}$10^{-14}$        &  -0.002 &      4&   -721  &  \\
 \hspace{3mm}$10^{-13}$        &  -0.003 &     -5&    831  &  \\[2mm]
\multicolumn{5}{l}{\em Other}\\
 \hspace{3mm}Area/Mass$=0$     &   -0.001 &      5&  -1122 &  \\
 \hspace{3mm}DE405 w/BIG-16    &   -0.048 &     84& -14160 &  \\
\hline
\end{tabular}
\end{center}
{\em Note}: For reference, the formal uncertainty in the $da/dt$ estimate is $0.100\times10^{-4}$ au/My and in $\zeta_{2135}$ is roughly 60,000 km.
\end{table}

\subsection{Gravitational Perturbers}\label{sec:pert}

The gravitational effects of the Sun, eight planets, the Moon and
Pluto are based on JPL's DE424 planetary ephemeris \citep{de424}. The
use of the older DE405 planetary ephemeris \citep{de405} leads to a
modest variation in the estimated $da/dt$ and the predicted
$\zeta_{2135}$ as indicated in Table~\ref{tab:models}.

When Bennu is near the Earth we modeled the gravitational
perturbation due to Earth oblateness. Table~\ref{tab:models} indicates
the effect of varying the distance within which the oblateness model
is included. We found that unless the effect was included whenever the
asteroid is closer than 0.3 au there is a modest but discernible
effect on the orbit determination and propagation. As a result we used
a 1 au cutoff as our baseline.

Perturbing asteroids were also included in the force model. Using DE424, we
developed mutually perturbed trajectories of the four largest
asteroids (1 Ceres, 2 Pallas, 4 Vesta and 10 Hygeia) and designated
this perturber model the CPVH small body ephemeris. We then
computed the orbits for the next 12 largest main belt asteroids, each of
which was perturbed only by DE424 and CPVH. The combination of these
12 additions with CPVH formed a perturber list of the 16 most massive
asteroids (based on current mass estimates), and we refer to this
perturber model as BIG-16. Finally, we added nine more asteroids,
which were selected according to an analysis of which perturbers could
most significantly influence the orbit of Bennu. The final
nine asteroid ephemerides, each perturbed by DE424 and BIG-16, were
combined with BIG-16 to form our final, baseline perturber set of 25
asteroids.

Table~\ref{tab:models} indicates the effect on the estimated value of
$da/dt$ due to changing the perturber model to either BIG-16 or
CPVH. In either case the effect is small and far less than the 0.5\% formal 
uncertainty. Table~\ref{tab:sbgm} lists
the assumed masses for each of the asteroid perturbers, as well as the
effect of deleting each one of them from the perturber model. From
this table one can see that, besides the very large contribution of 1
Ceres, 2 Pallas and 4 Vesta, only two other asteroids affect $da/dt$
at more than $0.1\sigma$, namely 6 Hebe and 7 Iris. Beyond CPVH, the
accumulation of smaller and smaller contributions tends toward a zero
mean. This is not surprising, and is a fortuitous result of the low aphelion distance (1.36 au)  of Bennu, which limits the perturbations of the main asteroid belt.

\begin{table}[t]
  \caption{Main belt asteroid perturbers, associated $GM$ values and the
    dynamical relevance of each as in the previous table.\label{tab:sbgm}}
\small
\begin{center}
\begin{tabular}{rrlr.rr}
\hline
\multicolumn{1}{c}{Count}&\multicolumn{1}{c}{IAU}&\multicolumn{1}{l}{Name}&\multicolumn{1}{c}{$GM$}&
 \multicolumn{1}{c}{ $\Delta da/dt$}&\multicolumn{1}{c}{$\Delta\xi_{2135}$} &\multicolumn{1}{c}{$\Delta\zeta_{2135}$} \\
       &   No.    &                 & 
       \multicolumn{1}{c}{(km$^3$/s$^2$)}& 
       \multicolumn{1}{c}{($10^{-4}$ au/My)} & 
       \multicolumn{1}{c}{(km)}& 
       \multicolumn{1}{c}{(km)} \\
\hline
\multicolumn{4}{l}{\em CPVH} &&\\
    1  &    1  &  Ceres           & 63.13$^a$   &  -0.118 &    298& -51351  \\
    2  &    2  &  Pallas          & 13.73$^b$   &  -0.109 &    295& -50754  \\
    3  &    4  &  Vesta           & 17.29$^c$   &   0.437 &   -893& 136194  \\
    4  &   10  &  Hygiea          &  5.78$^a$   &   0.002 &    -11&   1661  \\[2mm]
\multicolumn{4}{l}{\em Additions for BIG-16} &&\\
    5  &    3  &  Juno            &  1.82$^d$   &   0.002 &     -8&   1126 \\
    6  &    6  &  Hebe            &  0.93$^d$   &   0.019 &    -65&  10790 \\
    7  &    7  &  Iris            &  0.86$^d$   &  -0.014 &     57&  -9532 \\
    8  &   15  &  Eunomia         &  2.10$^d$   &  -0.001 &      1&   -260 \\
    9  &   16  &  Psyche          &  1.81$^d$   &  -0.005 &     16&  -2615 \\
   10  &   29  &  Amphitrite      &  0.86$^d$   &   0.000 &     -0&     30 \\
   11  &   52  &  Europa          &  1.59$^d$   &   0.001 &     -5&    643 \\
   12  &   65  &  Cybele          &  0.91$^d$   &   0.000 &      1&   -188 \\
   13  &   87  &  Sylvia          &  0.99$^d$   &   0.000 &     -1&     66 \\
   14  &   88  &  Thisbe          &  1.02$^d$   &  -0.007 &     19&  -3099 \\
   15  &  511  &  Davida          &  2.26$^d$   &   0.000 &      1&   -114 \\
   16  &  704  &  Interamnia      &  2.19$^d$   &   0.000 &      0&     -7 \\[2mm]
\multicolumn{4}{l}{\em Additions for Bennu} &&\\
   17  &   11  &  Parthenope      &  0.39$^d$   &  -0.006 &     18&  -3031    \\
   18  &   14  &  Irene           &  0.19$^d$   &  -0.002 &      6&  -1013    \\
   19  &   56  &  Melete          &  0.31$^d$   &  -0.002 &     -6&   1070    \\
   20  &   63  &  Ausonia         &  0.10$^d$   &  -0.002 &      4&   -731    \\
   21  &  135  &  Hertha          &  0.08$^d$   &   0.001 &     -6&    872    \\
   22  &  259  &  Aletheia        &  0.52$^d$   &   0.000 &      0&    -31    \\
   23  &  324  &  Bamberga        &  0.69$^d$   &  -0.001 &     -4&    628    \\
   24  &  419  &  Aurelia         &  0.12$^d$   &   0.002 &      2&   -259    \\
   25  &  532  &  Herculina       &  0.77$^d$   &   0.005 &    -13&   2200    \\
\hline
\end{tabular}
\end{center}
{\em Refs.}: $^a$\cite{bcm11}, $^b$\cite{konopliv_etal_2011}, $^c$\cite{russell_dawn}, $^d$\cite{carry_2012}.
\end{table}

\subsection{Relativity}\label{sec:rel}

\begin{figure}[t]
\begin{center}
\includegraphics[width=5in]{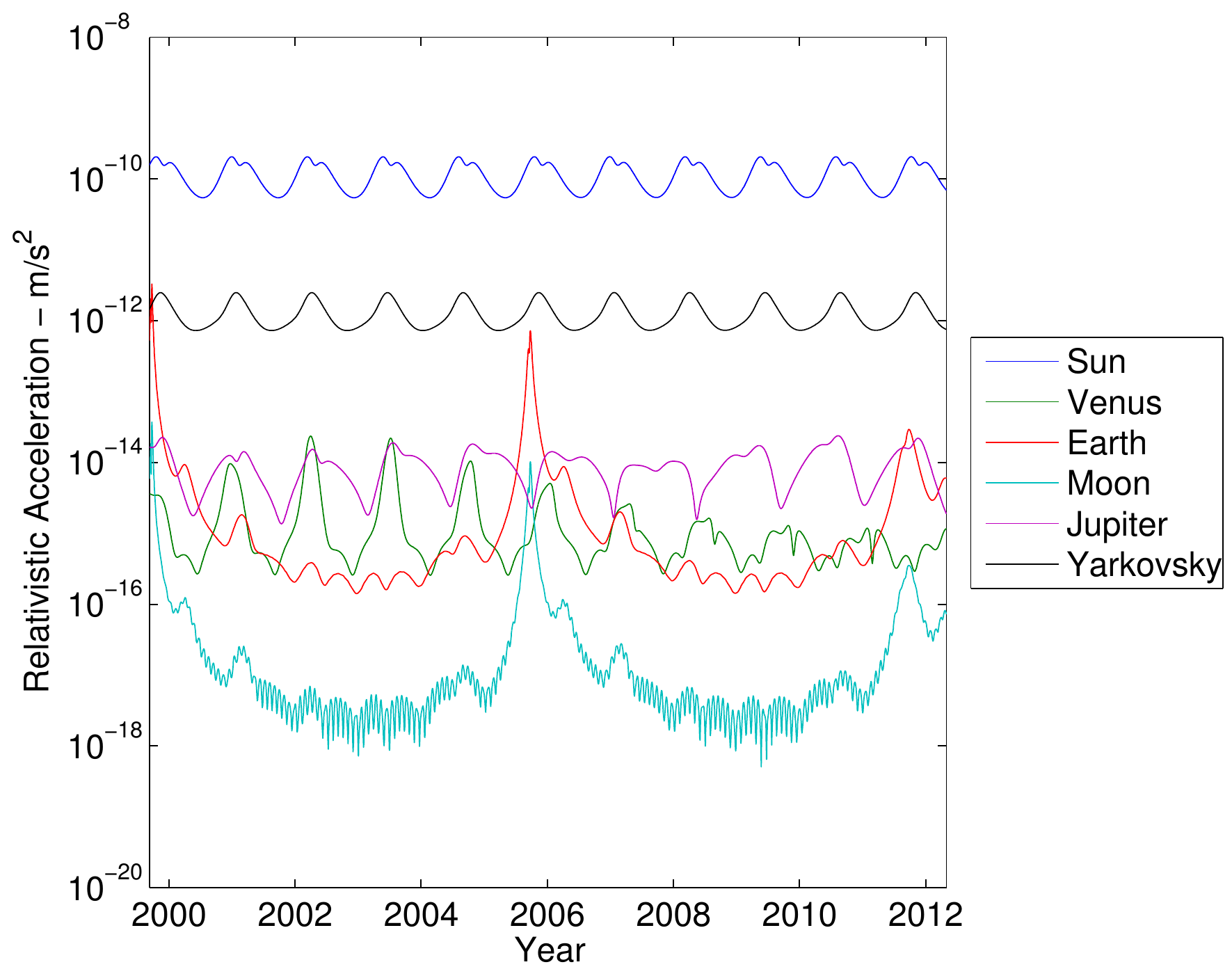}
\caption{Comparison of the Yarkovsky effect with the relativistic
  perturbations on Bennu. The magnitude of the respective
  accelerations is plotted.}
\label{fig:eih_rel}
\end{center}
\end{figure}

We used a full relativistic force model including the contribution of the Sun, the planets, and the Moon. More specifically, we used the Einstein-Infeld-Hoffman (EIH) approximation \citep{will_1993, moyer_2003, soffel_iau2000}. Table~\ref{tab:models} shows the variations in $da/dt$ and $\zeta_{2135}$ associated with different relativistic models. We found a 1.6\% difference in $da/dt$ with respect to the basic Sun-only Schwarzschild term \cite[Sec.~4]{damour}. This is only in small part due to the switch to the improved model for the Sun, as the contribution of some of the planets is not negligible. In particular, the Earth's relativistic terms are responsible for a 1.5\% ($\sim 3\sigma$) variation because of significant short range effects during Bennu Earth approaches in 1999 and 2005. Figure~\ref{fig:eih_rel} shows the main relativistic terms and compares them to the Yarkovsky perturbation. Clearly, the relativistic effects of the Sun are very important, about two orders of magnitude greater than Yarkovsky, though it matters little whether the Schwarzschild or EIH approximation is used. The Earth's relativistic terms are at the same level as Yarkovsky during the Earth encounters in 1999 and 2005. At other times, Jupiter and Venus perturbations are generally more significant, although even the lunar term can briefly exceed them during close Earth encounters.

\begin{figure}[t]
\begin{center}
\includegraphics[width=5in]{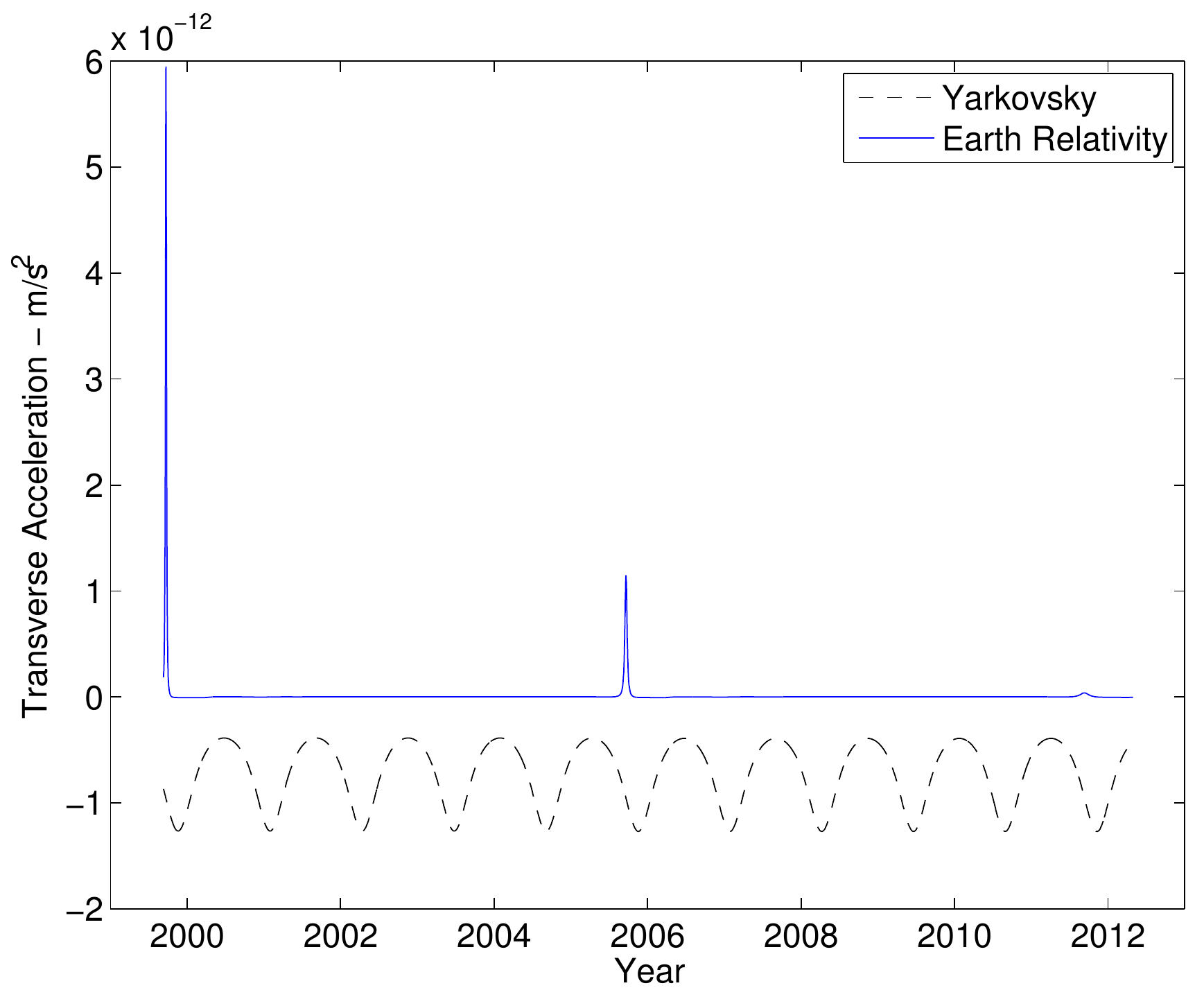}
\caption{Comparison of the transverse perturbations from the Yarkovsky
  effect and the relativistic component due to the Earth. Earth relativity
  can be significantly greater in magnitude than the Yarkovsky effect during
  Earth encounters, and thus provides a statistically significant
  change in semimajor axis that must be accounted for in the dynamical
  model.}
\label{fig:yarko_eih}
\end{center}
\end{figure}

The Yarkovsky effect is primarily a transverse acceleration and thus the transverse component of the relativistic perturbations can alias as Yarkovsky if not properly modeled. Figure~\ref{fig:yarko_eih} depicts how the transverse component of Earth relativistic perturbation during the 1999 close approach is several times greater than the transverse acceleration associated with the Yarkovsky effect. Because the modeled semimajor axis drift is an integral of the two curves in Fig.~\ref{fig:yarko_eih}, neglecting Earth relativity leads to significant errors.

Table~\ref{tab:models} indicates that Earth's relativity term is the most significant factor among all of those considered, at least on longer timescales as indicated by $\Delta\zeta_{2135}$. On shorter timescales, i.e., during the fitspan, Table~\ref{tab:sbgm} reveals that the perturbation of Vesta leads to a greater change in $\Delta da/dt$ than Earth relativity, although Earth relativity is still more important than Vesta for longer integrations. The uncertainty in both of these perturbations is negligible.

\subsection{Outlier Treatment}\label{sec:outlier}

The selection of outliers has a statistically significant effect on
the orbital prediction. To explore this sensitivity, we have generated
for comparison several orbital solutions with a variety of automatic
outlier rejection parameter settings. These are summarized in
Table~\ref{tab:models}, which lists the $\chi_{\rm{rej}}$ parameter
value used in the algorithm described by \citet{carpino_etal03}. The outlier rejection threshold $\chi_{\rm{rej}}$ is similar to the sigma level at which outlier rejection takes place, but the algorithm is more sophisticated than simple sigma clipping. The
number of observations deleted in the various cases is also
tabulated. The importance of careful attention to statistical outliers
is indicated by the fact that the solutions are seen to progress
steadily towards solution 87 as progressively more stringent requirements
are placed on the outlier selection. In terms of estimated $da/dt$,
the most inclusive approach to outliers falls about $0.5\sigma$ from
the solution 87 estimate. However, even a cursory inspection of the data
indicates that numerous spurious points remain in the fit for that
solution. While the manual outlier rejections in solution 87 are more
aggressive than even the most stringent automatic selections (e.g.,
$\chi_{\rm{rej}}=1$), the separation between these two solutions is
slight, and both are very well constrained by the observational data,
with 478 and 519 optical observations, respectively. Most of the
movement in the orbital predictions due to outlier treatment can be
traced to a small handful of observatories with significantly biased
observations. In general, the manual approach deletes more
observations because it often removes the entire contribution from a
problematic observatory, rather than only those that are clearly
discordant with the bulk of the data.

\subsection{Numerics and Software Validation}\label{sec:num_val}

\citet{giorgini_1950da} show that numerical integration errors are not significant for the case of the year 2880 potential impact of 1950 DA. We reach the same conclusion for Bennu by varying the integration error tolerance used in our software. Table~\ref{tab:models} shows that the estimated value of $da/dt$ is not materially affected by integrator tolerance values $\le 10^{-14}$.

All of the results in this paper are based on the outputs of the JPL orbit determination and propagation software package. We have verified our primary JPL results by careful cross-referencing with comparable results obtained with the OrbFit orbit determination and integration package\footnote{See http://adams.dm.unipi.it/orbfit; we used the OrbFit version 4.3, which is still in beta testing.}. We compared the orbital solutions, the sensitivity to different settings of the dynamical model, and the orbit propagation, and we found that these two independent software packages reproduce each other's results very well. Indeed, the comparison with the OrbFit software package revealed to us the critical importance of the Earth general relativity terms in the dynamical model. After resolving modeling discrepancies, we found that the determinations of $A_T$ from the orbital fits was consistent to better than 0.1\%, corresponding to $<0.2\sigma$ in $da/dt$. We are therefore confident that our findings are not corrupted by software bugs.

%
%

\section{Mass, Bulk Density and Implications}\label{sec:rho}

The linear and nonlinear Yarkovsky models both require the asteroid bulk density $\rho$, which was initially unknown. However, since all other parameters in the model are independently known, we can estimate this quantity. We used the linear model to compute JPL solution 86, with an associated bulk density estimate of $1314$ kg/m$^3$. Similarly, we used the more accurate nonlinear model to obtain JPL solution 85 (Table~\ref{tab:nominal}), which includes an associated bulk density estimate $\rho=1181$ kg/m$^3$. The discrepancy between the two models is a combination of factors, but overall implies that the linear model overestimates the transverse Yarkovsky force by about $11\%$ and thus the estimated value of $\rho$ is increased to maintain the required mean $da/dt$. Of particular importance is the oblateness of the Bennu shape model. This flattening leads to a diminished cross-sectional area, which tends to reduce the energy input and thereby reduce the thermal recoil acceleration in the nonlinear model. According to the theory of \citet{vok_1998} this should account for a 5--10\% error. Additionally, the linearization of the heat transfer problem tends to slightly increase the thermal emissions \citep{capek_belgrade}, which readily accounts for the remaining discrepancy.

The uncertainty in the bulk density estimate is a complex story due to the numerous parameters that are used in formulating the estimate. The formal uncertainty that is obtained directly from the orbit determination (Table~\ref{tab:nominal}) captures only the $0.5\%$ uncertainty in the semimajor axis drift, and does not account for the more significant sources of uncertainty outlined in Table~\ref{tab:rq36_props}. The final column of that table indicates how the associated parameter uncertainty maps into the bulk density uncertainty, from which we conclude that the uncertainty in thermal inertia and asteroid size dominate over other error sources. 

As described in Sec.~\ref{sec:yarko}, for a sphere we are sensitive to the product $\rho D$ and so the density estimate varies inversely with the asteroid size, in contrast to other density estimates that are derived from the asteroid volume. For a non-spherical shape the Yarkovsky acceleration actually depends on the quotient of the radiative cross-sectional area and the volume $A/V$, rather than $1/D$, and yet the contribution to bulk density is still linear.

\begin{figure}[t]
\begin{center}
\includegraphics[width=5in]{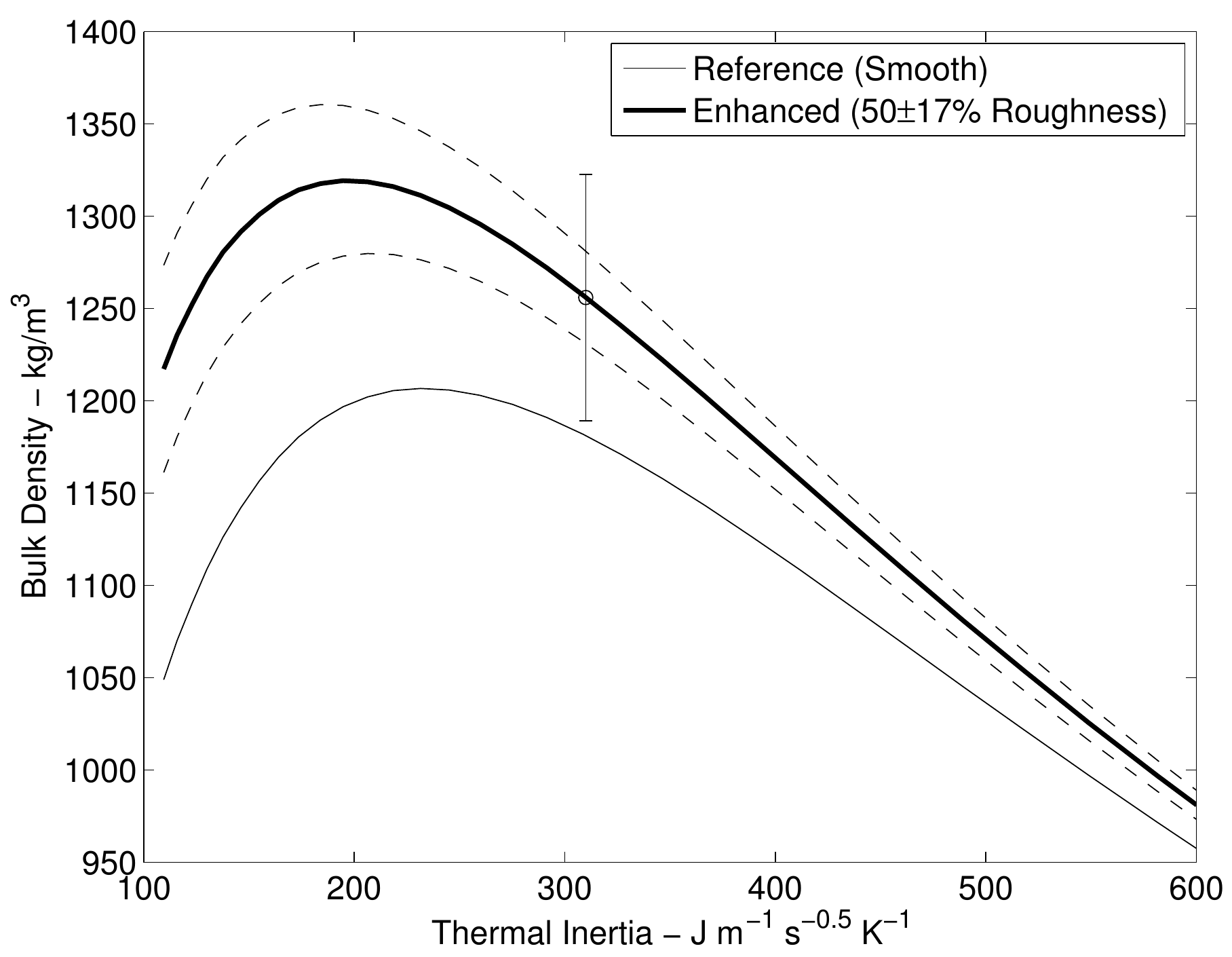}
\caption{The bulk density estimate for Bennu depends nonlinearly on the estimated thermal inertia $\Gamma$. Neglecting surface roughness, we obtain a bulk density estimate of 1180 kg/m$^3$. However, taking into account the assumed Yarkovsky enhancement from roughness, as well as uncertainties in obliquity, diameter and thermal inertia, we obtain $1260\pm70$ kg/m$^3$ as depicted here.}
\label{fig:rho}
\end{center}
\end{figure}

In contrast, the bulk density dependence on thermal inertia is markedly nonlinear (Fig.~\ref{fig:rho}). The thermal inertia of Bennu is $\Gamma=310\pm70\; \rm{J\,m^{-1}\,s^{-0.5}\,K^{-1}}$ \citep{emery_bennu_2014}. This value is derived from analysis of a suite of observations of thermal flux, consisting of 8--20 $\mu$m spectra of opposite hemispheres and photometry at 3.6, 4.5, 5.8, 8.0, 16, and 22 $\mu$m of 10 different longitudes using the Spitzer Space Telescope. The thermophysical modeling that results in this thermal inertia estimate incorporates the detailed shape and spin information derived from radar imaging and visible light curve photometry, and explicitly includes the effects of macroscopic surface roughness. 

Previous estimates of Bennu's thermal inertia \citep{emery_lpsc_2010, emery_dps_2012, muller_2012_rq36} are somewhat higher ($\sim600\; \rm{J\,m^{-1}\,s^{-0.5}\,K^{-1}}$), and have led to correspondingly lower bulk density estimates \citep[e.g., 970 kg/m$^3$,][]{chesley_acm2012}. There are two primary reasons for the different thermal inertia values. First, the earlier studies used only a subset of the Spitzer data, namely the spectra. Those spectra are noisy, making it difficult to scale the different segments of the full spectra relative to each other. Different scale values affect the best-fit model surface temperature distribution, and therefore the derived thermal inertia. In contrast, \citet{emery_bennu_2014} include the large set of photometric data, which have much higher signal-to-noise than the spectral data, leading to results that are both more accurate and have significantly smaller uncertainties. Note that if the uncertainties in scale factors are included in the uncertainty estimates from the spectral data, the error bars overlap the \citet{emery_bennu_2014} estimate given above. Second, the earlier estimates assumed a spherical shape for Bennu. However, Bennu is actually fairly oblate \citep{nolan_etal_2013}. The oblateness causes surface facets to be tilted farther away from the Sun as compared to a sphere. The models assuming spherical shape compensate for the more direct viewing geometry with a lower thermal inertia. For these reasons, we rely here on the updated thermal inertia from \citet{emery_bennu_2014}.

\begin{figure}[t]
\begin{center}
\includegraphics[width=5in]{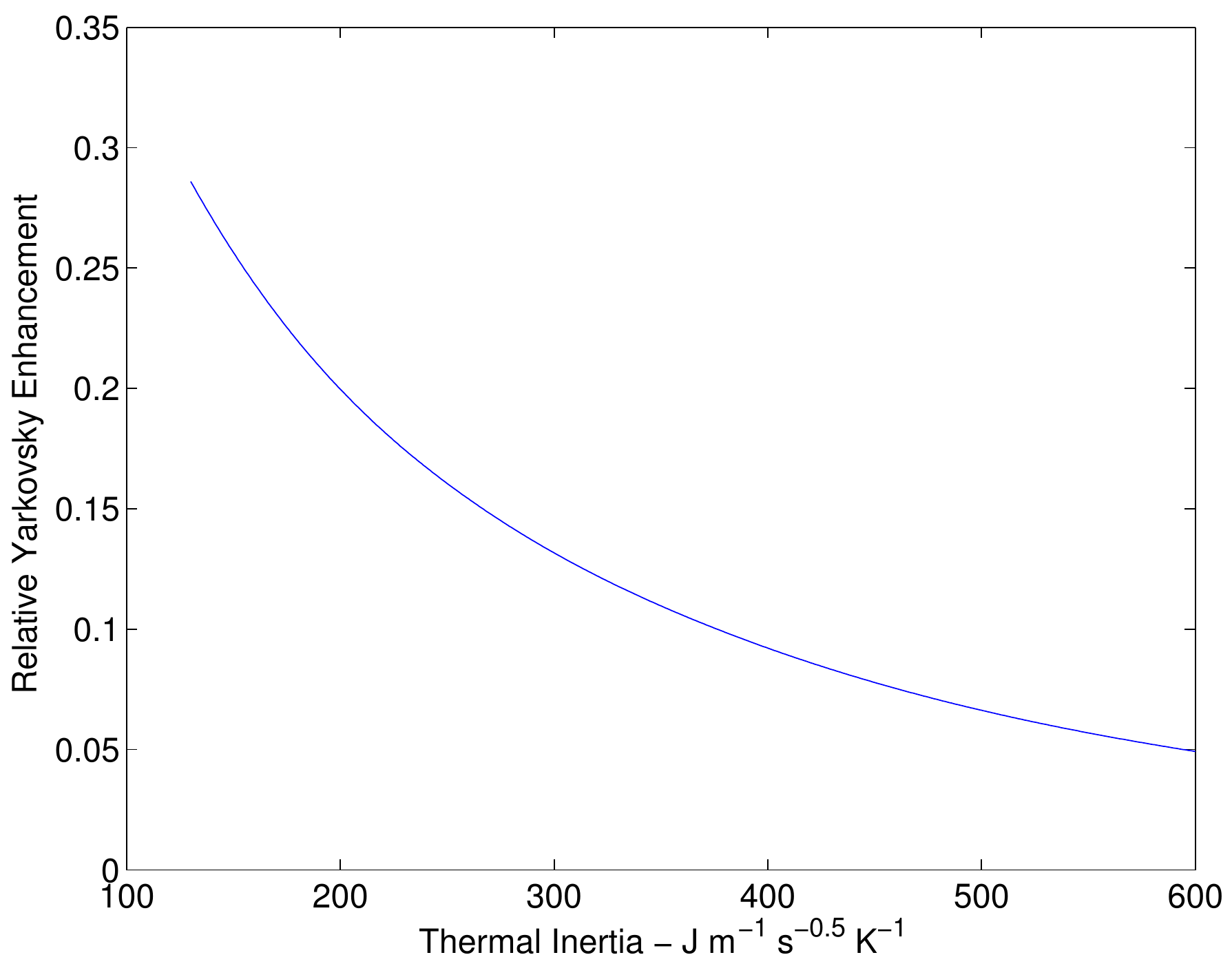}
\caption{Yarkovsky effect enhancement due to 100\% surface roughness as a function of thermal inertia.}
\label{fig:enhance}
\end{center}
\end{figure}

The effects of surface roughness on Bennu are not incorporated into the $\rho$ estimates so far, and yet \citet{rozitis_green_yarko} used a sophisticated thermophysical model to show that the thermal effects of surface roughness always tend to increase the Yarkovsky effect. For the Bennu shape model with the roughest surface model the increase in $\rho$ is 12.7\% at the nominal thermal inertia (see Fig.~\ref{fig:enhance}), pointing to roughness as the dominant source of uncertainty. 

Although we have no rigorous estimates of Bennu's roughness, it is unlikely to be either remarkably smooth or extremely rough. The thermal inertia from \citet{emery_bennu_2014} is somewhat lower than that derived for (25143) Itokawa \citep{muller_2005_itokawa}, which suggests that Bennu could have a somewhat smoother surface texture than Itokawa. Elsewhere, \citet{nolan_etal_2013} find that the radar circular polarization ratio, which is a proxy for near-surface roughness at the scale of the radar wavelength (12.6 and 3.5 cm), indicates a relatively smooth surface compared to other bodies that are not particularly rough. In particular, they find Bennu has significantly lower polarization ratios than Itokawa at both wavelengths and conclude that Bennu is likely smoother than Itokawa. However, \citet{nolan_etal_2013} did identify a boulder on Bennu with a size of 10--20 m, suggesting the presence of smaller boulders below the resolution limit of 7.5 m and a surface that is not perfectly smooth. In the absence of reliable estimates, we assume that the roughness is $50\pm17$\%, which covers the full range 0--100\% at $3\sigma$. This yields the ``Enhanced'' curve in Fig.~\ref{fig:rho}. Inflating the reference value $\rho=1181$ kg/m$^3$ by 50\% of the 12.7\% enhancement yields our best estimate of 1255 kg/m$^3$.

To develop a comprehensive estimate of the uncertainty in the presence of the nonlinearity evident in Fig.~\ref{fig:rho} we take a Monte Carlo approach. We sample $\Gamma$, $A/V$ and $\gamma$ according to the normal distributions given by Table~\ref{tab:rq36_props}. We obtain the Yarkovsky enhancement for each case by sampling a roughness from $50\pm17$\% and using it to scale the 100\%-rough enhancement (from Fig.~\ref{fig:enhance}) at the sampled thermal inertia. This leads to our final bulk density estimate of $\rho = 1260 \pm 70$ kg/m$^3$. The associated mass, $GM$ and area-to-mass ratio values are listed in Table~\ref{tab:mass}.

\begin{table}[t]
\caption{Bennu bulk density and related quantities with $1\sigma$ uncertainties.\label{tab:mass}}
\small
\begin{center}
\begin{tabular}{ll}
\hline
Bulk density $\rho$ ($\mathrm{kg/m^3}$) 		& $1260\pm70$ \\[1mm]
Mass $M$ ($10^{10}\,\rm{kg}$)            		& $7.8\pm0.9$ \\[1mm]
$GM$ ($\mathrm{m^3/s^2}$)            			& $5.2\pm0.6$ \\[1mm]
Area-to-mass ratio ($10^{-6}\,\rm{m^2/kg}$)		& $2.4\pm0.1$ \\[1mm]
\hline         
\end{tabular}
\end{center}
\end{table}

We assume that the Yarkovsky effect is the only significant source of nongravitational acceleration on Bennu, and in particular we do not account for the possibility of outgassing, which would corrupt our bulk density estimate if it were significant. If the direction of any hypothetical outgassing is skewed towards the evening terminator, which might be expected as the diurnal thermal wave penetrates to release buried volatiles, then it would combine to increase the magnitude of the transverse acceleration on the asteroid. In this case, the bulk density estimate should be increased to account for outgassing. Conversely, if the outgassing tends to cancel the transverse thermal recoil acceleration then our bulk density estimate will be an overestimate.

3200 Phaethon, a B-type asteroid like Bennu, has long been identified as the parent body of the Geminid meteor shower \citep{whipple_geminids}. This suggests that at least some objects of this taxonomic type have the possibility of shedding material, possibly as fine-grained material entrained in gasses released by the sublimation of volatiles. However, the only report of possible dust release on Phaethon took place at a heliocentric distance of 0.14 au, and it has been suggested that the Geminids are shed from Phaethon due to fracturing associated with fatigue from thermal cycling and the decomposition of hydrated minerals \citep{jewitt_li_2010}. There is no evidence that Bennu is shedding material and, with $q\simeq 0.9\,\rm{au}$, solar heating is markedly lower than that experienced by Phaethon. Therefore, we do not consider it likely that outgassing is significantly affecting our results.

Is the estimated $GM$ in Table~\ref{tab:mass} consistent with loose material on the equator being gravitationally bound to the surface? \citet{nolan_etal_2013} report that the maximum equatorial diameter is 565 m. From this and the known spin rate we find a lower bound of $GM=3.7\,\rm{m^3/s^2}$ if we assume that the gravitational attraction exceeds the centrifugal acceleration. This is a reasonable expectation because, as discussed above, we consider it likely that Bennu's surface is dominated by cm-scale and smaller regolith. However, it is difficult to rule out the possibility that some regions along the equator are devoid of loose material, or that induration or cohesion provides sufficient binding to keep material on the surface that would otherwise depart. Even so, the assumption that material is gravitationally bound to the surface would imply $\rho > 890\,\rm{kg/m^3}$ which is satisfied here with a high degree of confidence.

One can compute the macroporosity, $P=1-\rho/\rho_M$, of an asteroid if the bulk densities of the body $\rho$ and the appropriate meteorite analog $\rho_M$ are known. Bennu has been identified as a B-type asteroid, and CI and CM carbonaceous chondrite meteorite samples provide the best spectral match \citep{clark_etal_2011}. \citet{consolmagno_etal_cde_2008} report that CM meteorite samples have average bulk densities  of $2130\pm 190\,\rm{kg/m^3}$, which, taken together with our asteroid bulk density estimate and uncertainty from Table~\ref{tab:mass}, suggests $P$ in the range 30--50\%. For CI meteorites the data are fewer and less conclusive, with different measurement techniques leading to sample bulk densities similar to those of CM meteorites or as low as $1600\,\rm{kg/m^3}$, which would allow $P$ to be as low as 20\% \citep{consolmagno_etal_cde_2008}. Overall, our judgement is that the macroporosity of Bennu is likely to be in the range $40\pm10$\%, but could be as low as 20\%.

Bennu's estimated bulk density is comparable to values obtained for other low-albedo asteroids, from large asteroids in the main asteroid belt to smaller asteroids in the inner solar system. The average C-type asteroid in the main belt, according to estimates derived from the gravitational perturbations on the planets, predominantly Mars, is $\rho=1290\pm 60\,\rm{kg/m^3}$ \citep{de405}. This estimate is biased toward asteroids with diameters much larger than $100$ km, which contain the majority of the mass among C-types. However, a flyby of the 53-km, C-type asteroid (253) Mathilde by the NEAR spacecraft also yielded a similar density $\rho=1300\pm 200\,\rm{kg/m^3}$ \citep{yeomans_etal_97}. Furthermore, \citet{marchis_2008_circ, marchis_2008_ecc} report densities of several C-complex binary asteroids in the main belt. The summary given by \citet[Table 8]{marchis_2008_circ} suggests that the density distribution of large, C-complex binaries is $\rho=1100\pm 300\,\rm{kg/m^3}$. Among the near-Earth asteroid population, \citet{shepard_2006ce26} report that the low-albedo binary system (276049) 2002 CE$_{26}$ has a 3.5-km primary with bulk density $900^{+500}_{-400}\,\rm{kg/m^3}$, which is comparable within the error bars to that of Bennu.

What do Bennu's density and porosity tell us?  To say anything useful here, we need to put Bennu into context. Our best estimates suggest Bennu is a fragment of a larger body that experienced a collision \citep{campins_rq36_2010, walsh_newpolana}. Similarly, the large multiple systems examined by \citet{marchis_2008_circ, marchis_2008_ecc} were presumably formed by large collision events \cite[e.g.,][]{durda_2004}. One would expect these smashed up target worlds, with porosity added by the fragmentation and ejecta re-assembly process, to have low bulk densities in comparison to their meteorite analogs, and that the smallest bodies should tend to still lower density due to self-gravitation and compaction on the large bodies \citep{baer_chesley_2008, bcm11}. However, this size dependence seems to vanish at sizes below roughly 250--300 km, below which no obvious size trend exists in macroporosity, an observation reinforced by Bennu. With this background, we argue that Bennu's porosity was produced by a similar mechanism, consistent with our hypothesis that void space and porosity were added into Bennu by its formation and/or by post-formation processes. Taken together, these arguments allow us to infer that Bennu has a heavily fractured or shattered internal structure combined with a substantial porosity. These characteristics fit the definition of a rubble pile asteroid provided by \citet{richardson_etal_2002}.

\section{Earth Close Approaches}\label{sec:future}
%


The deterministic prediction interval for the trajectory of Bennu extends for 481 years, from 1654 to 2135. Earth close
approaches within 0.05 au during this time interval are listed in
Table~\ref{tab:det_ca}. Close encounters outside of this interval have
encounter time uncertainties well in excess of a day. The closest
approach in this interval is the nominally sub-lunar distance
encounter in 2135. This deep close approach leads to strong scattering
of nearby orbits, and so the subsequent impact hazard can only be
explored through statistical means.

Figure~\ref{fig:elem_hist} shows the time history of Bennu's
orbital elements from 2000 to 2136. There are variations of a few
percent due to Earth close approaches, especially in 2135. As the
Yarkovsky induced orbital drift depends on the osculating orbital
elements \citep{farnocchia_yarkolist_2013}, there are also commensurable
variations in the $da/dt$ evolution (see Fig.~\ref{fig:dadt_hist}).

Table~\ref{tab:models} details the effect of various differing models on the $b$-plane coordinates $(\xi_{2135}, \zeta_{2135})$ of the close approach at the last reliably predicted Earth encounter for Bennu, which takes place in 2135. The $b$-plane is oriented normal to the inbound hyperbolic approach asymptote and is frequently used in encounter analysis. The $(\xi, \zeta)$ coordinates on the $b$-plane are oriented such that the projected heliocentric velocity of the planet is coincident with the $-\zeta$-axis. In this frame the $\zeta$ coordinate indicates how much the asteroid is early ($\zeta<0$) or late ($\zeta>0$) for the minimum possible distance encounter. In absolute value, the $\xi$ coordinate reveals the so-called Minimum Orbital Intersection Distance (MOID), which is the minimum possible encounter distance that the asteroid can attain assuming only changes to the timing of the asteroid encounter. For a more extensive discussion  of these coordinates see \citet{opik_valsecchi} and references therein. In Table~\ref{tab:models}, the tabulated $da/dt$ differences are indicative of the importance of the effect on the 1999--2012 time frame of the observation set, while $\zeta_{2135}$ provides an indication of how important the term is for the much longer integration from 2011 to 2135.

\begin{table}[t]
\caption{Bennu Deterministic Earth Approaches Closer than 0.05 AU (JPL solution 76).\label{tab:det_ca}}
\small
\begin{center}
\begin{tabular}{ccccrr}
\hline
   Date (TDB)     & Nom. Dist.&$3\sigma$ Min.&$3\sigma$ Max.& $1\sigma$ Time Uncert.& $1\sigma$ $\zeta$ Uncert.     \\
                 &  (AU)     &    (AU)      & (AU)         &  (s) &   (km)   \\
\hline         
1654 Sep 17.89194 & 0.022033 & 0.010586 & 0.035240 &  49590   &    816253     \\
1788 Sep 20.56364 & 0.009771 & 0.009304 & 0.010254 &   2237   &     32739     \\
1848 Sep 21.91904 & 0.007915 & 0.007892 & 0.007938 &    105   &      1573     \\
1911 Sep 22.88762 & 0.014178 & 0.014177 & 0.014179 &      2.4  &        38     \\
1970 Sep 27.10790 & 0.021403 & 0.021403 & 0.021403 &      1.6  &        16     \\
1999 Sep 22.76422 & 0.014686 & 0.014686 & 0.014686 &   $\ll1$  &         1.1   \\
2005 Sep 20.44528 & 0.033130 & 0.033130 & 0.033130 &   $\ll1$  &         1.2   \\
2054 Sep 30.04163 & 0.039299 & 0.039299 & 0.039299 &      1.6  &        11     \\
2060 Sep 23.02530 & 0.005008 & 0.005008 & 0.005008 &      1.0  &        15     \\
2080 Sep 22.02378 & 0.015560 & 0.015427 & 0.015693 &    360   &      7318     \\
2135 Sep 25.40942 & 0.002009 & 0.000819 & 0.003549 &   4746  &     79674     \\
\hline
\end{tabular}
\end{center}
\end{table}

\subsection{Impact Hazard Assessment}
\begin{figure}[t]
\begin{center}
\includegraphics[width=5in]{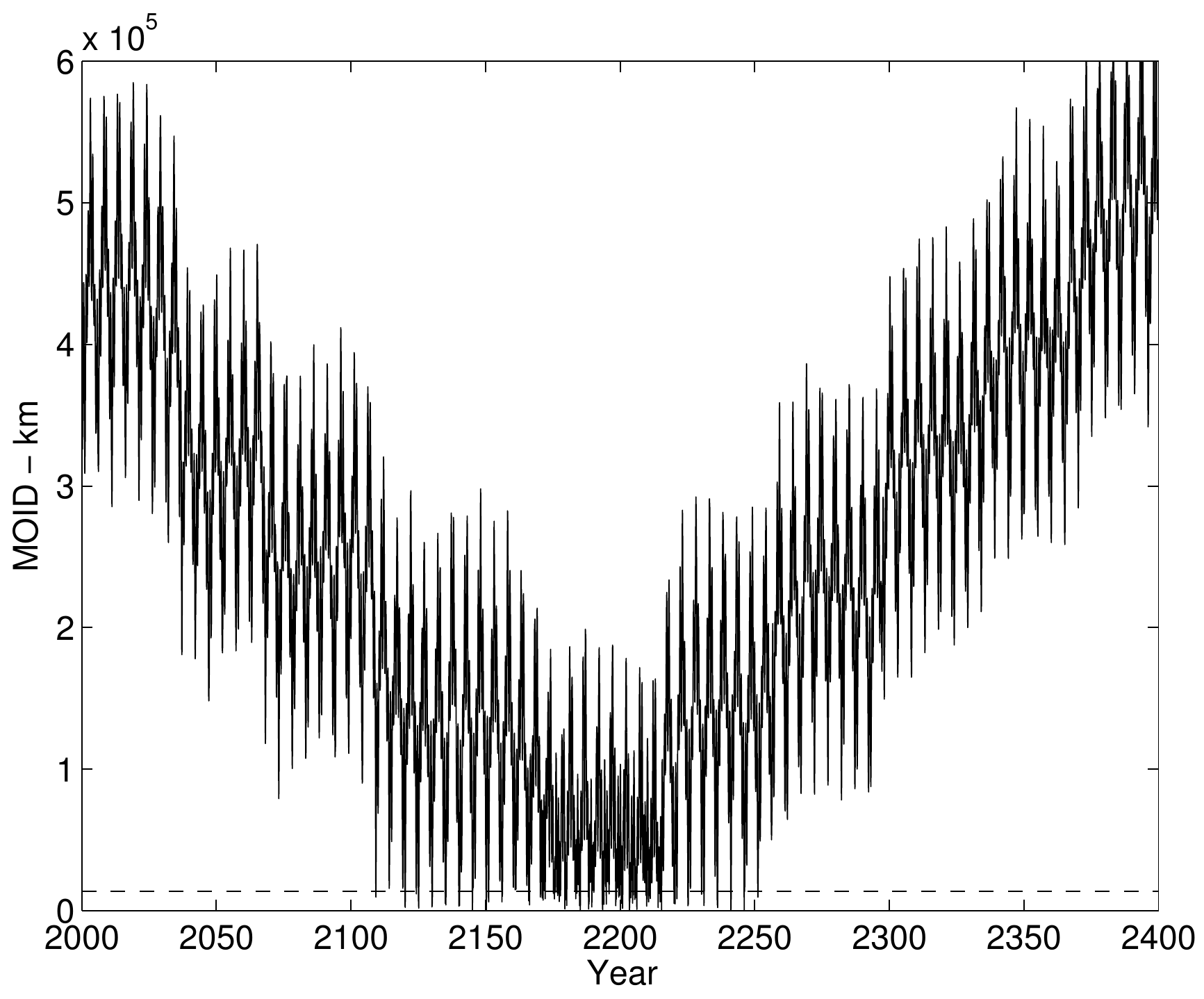}
\caption{Time history of the MOID. The Earth impact cross-section (2.1 Earth radii) is
marked by the dashed line.}
\label{fig:moid}
\end{center}
\end{figure}

The geometry of Bennu's orbit allows deep close approaches to the Earth, which require a careful assessment of the associated potential collision hazard. Figure~\ref{fig:moid} shows the dependence on time of the Minimum Orbit Intersection Distance \citep[MOID, see, e.g.,][]{gronchi_moid}. According to the secular evolution, the MOID reaches its minimum near the end of the next century while short periodic perturbations make it cross the Earth impact cross section threshold at different epochs from 2100 to 2300, which is therefore the time period for which we must analyze possible close approaches. This objective is similar to that discussed by \citet{milani_etal_rq36}, however we bring new analysis tools to bear on the problem and we have the benefit of crucial astrometric data not available in \citeyear{milani_etal_rq36}. We recall that \citet{milani_etal_rq36} based much of their analysis on the variability of the 2080 encounter circumstances, finding that, for the observational data then available, this was the last encounter that was well constrained, and after which chaotic scattering made linear analysis infeasible. With the current data set, future encounter uncertainties remain modest until after 2135 (Table~\ref{tab:det_ca}), and nonlinear analysis techniques are necessary for subsequent encounters. Thus the 2135 encounter is the central focus in our current impact hazard assessment.

\begin{figure}[t]
\begin{center}
\includegraphics[width=2in]{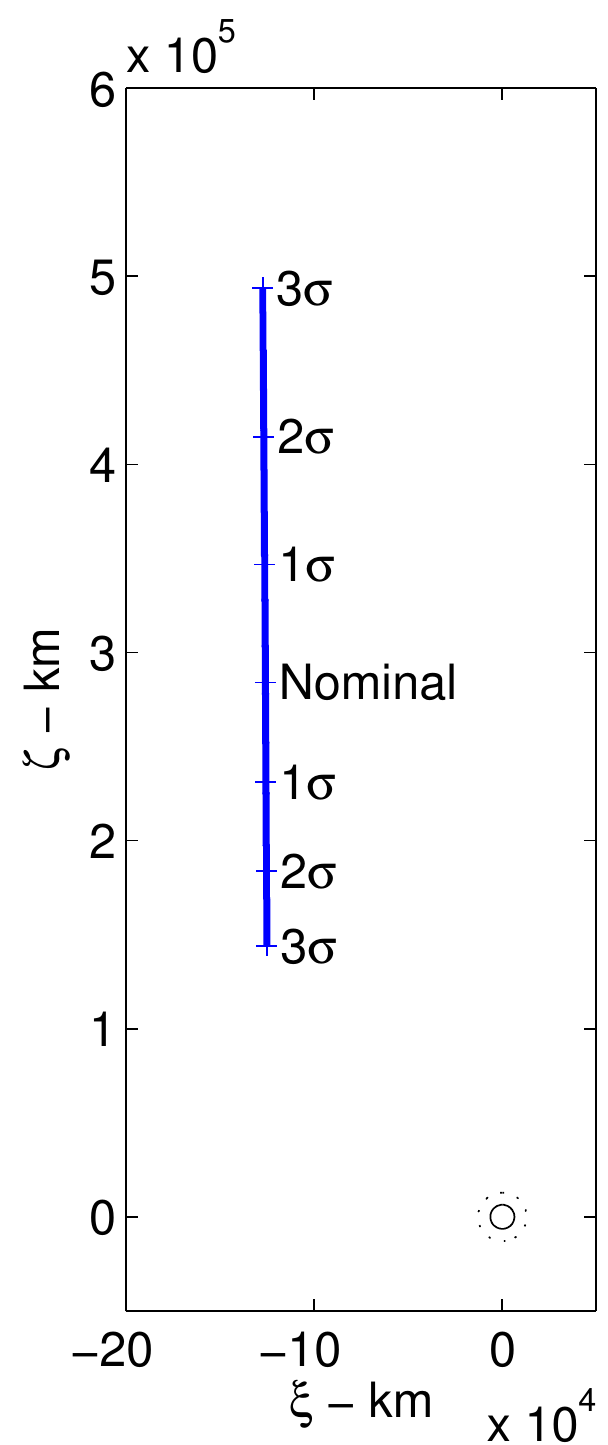}
\caption{Uncertainty region on the $b$-plane of the 2135
encounter. The Earth is plotted to scale at the origin, and the impact
cross-section is indicated by the dots.}
\label{fig:bplane}
\end{center}
\end{figure}

We performed a
Monte Carlo sampling \citep{chodas_yeomans99_girdwood} in the
7-dimensional space of initial conditions and bulk
density. Figure~\ref{fig:bplane} shows the distribution of the Monte
Carlo samples on the 2135 $b$-plane. The $b$-plane plot depicts the geocentric locations of the incoming hyperbolic asymptote of the Monte Carlo samples on the plane orthogonal to the asymptote, indicating the distance and direction of the closest approach point of a fictitious unperturbed trajectory \citep[see, e.g.,][]{opik_valsecchi}. The linear mapping of the
uncertainty region is a poor approximation as we can see from the
asymmetry of the distribution. As expected, the uncertainty region gets
stretched along $\zeta$, which reflects time of arrival variation and
is thus related to the along-track direction.

By propagating the Monte Carlo samples through year 2250 we can
determine the Virtual Impactors (VIs), i.e., the Virtual Asteroids
(VAs) compatible with the orbital uncertainty corresponding to an
impacting trajectory. The positions of the VIs in the 2135 $b$-plane
define the 2135 keyholes, which are the coordinates on the $b$-plane corresponding to a subsequent impact \citep{C99}. On the $b$-plane of a given
post-2135 encounter we can interpolate among nearby samples to
identify the minimum possible future encounter distance. When this minimum
distance is smaller than the Earth radius, the keyhole width is
obtained by mapping the chord corresponding to the intersection
between the line of variations and the impact cross section back to the 2135
$b$-plane. This procedure allows us to develop a map of the keyholes in the $b$-plane. For Bennu we found about 200 keyholes in the 2135 $b$-plane with widths ranging from 1.6 m to 54 km.

\begin{figure}[t]
\begin{center}
\includegraphics[width=5in]{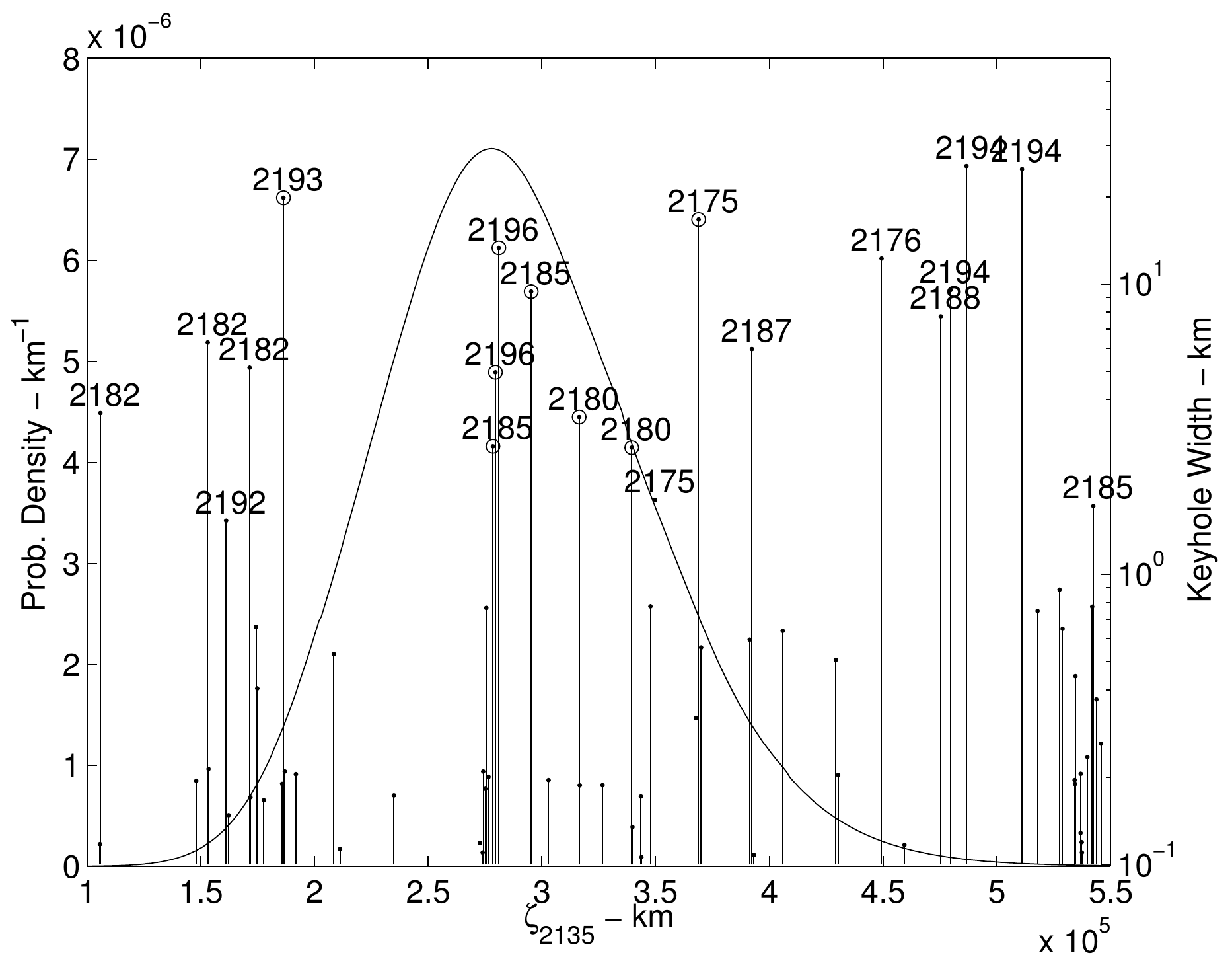}
\caption{A map of the Bennu impact keyholes on the 2135
$b$-plane. The probability density is given by the curve with the left
ordinate, and the keyholes are indicated by the vertical lines at
their respective $\zeta_{2135}$ positions with their widths given on
the right ordinate. For clarity only keyholes wider than 1 km are
labeled with the year of impact and only keyholes $> 100$ m in width
are depicted. Potential impacts with impact probability greater than
$10^{-5}$ are marked with a circle at the top of the vertical bar.}
\label{fig:keyhole}
\end{center}
\end{figure}

Figure~\ref{fig:keyhole} shows the probability density function (PDF)
of $\zeta_{2135}$ resulting from the Monte Carlo sampling. As already
noted, the linear approximation is not valid in this case, and so the PDF is distinctly non-gaussian. The figure also reveals the keyhole map in $\zeta_{2135}$, where the 
vertical bars correspond to the keyholes $>100$ m in width and the
height of the bars is proportional to the width. For a given keyhole
the impact probability (IP) is simply the product of
the PDF and the keyhole width. For each of the 78 keyholes larger than
$100$ m and with an IP $>10^{-10}$, Table~\ref{tab:keyhole} reports
the impact year, the keyhole width, the impact probability, and the
associated Palermo Scale \citep{palermo_scale}. The cumulative IP is
$3.7\times 10^{-4}$ and the cumulative Palermo Scale is -1.70. There are eight keyholes corresponding to an IP
larger than $10^{-5}$. Among these, the year 2196 has the highest IP, $1.3\times 10^{-4}$, which arises primarily from two separate but nearby keyholes.

\begin{figure}[t]
\begin{center}
\includegraphics[width=5in]{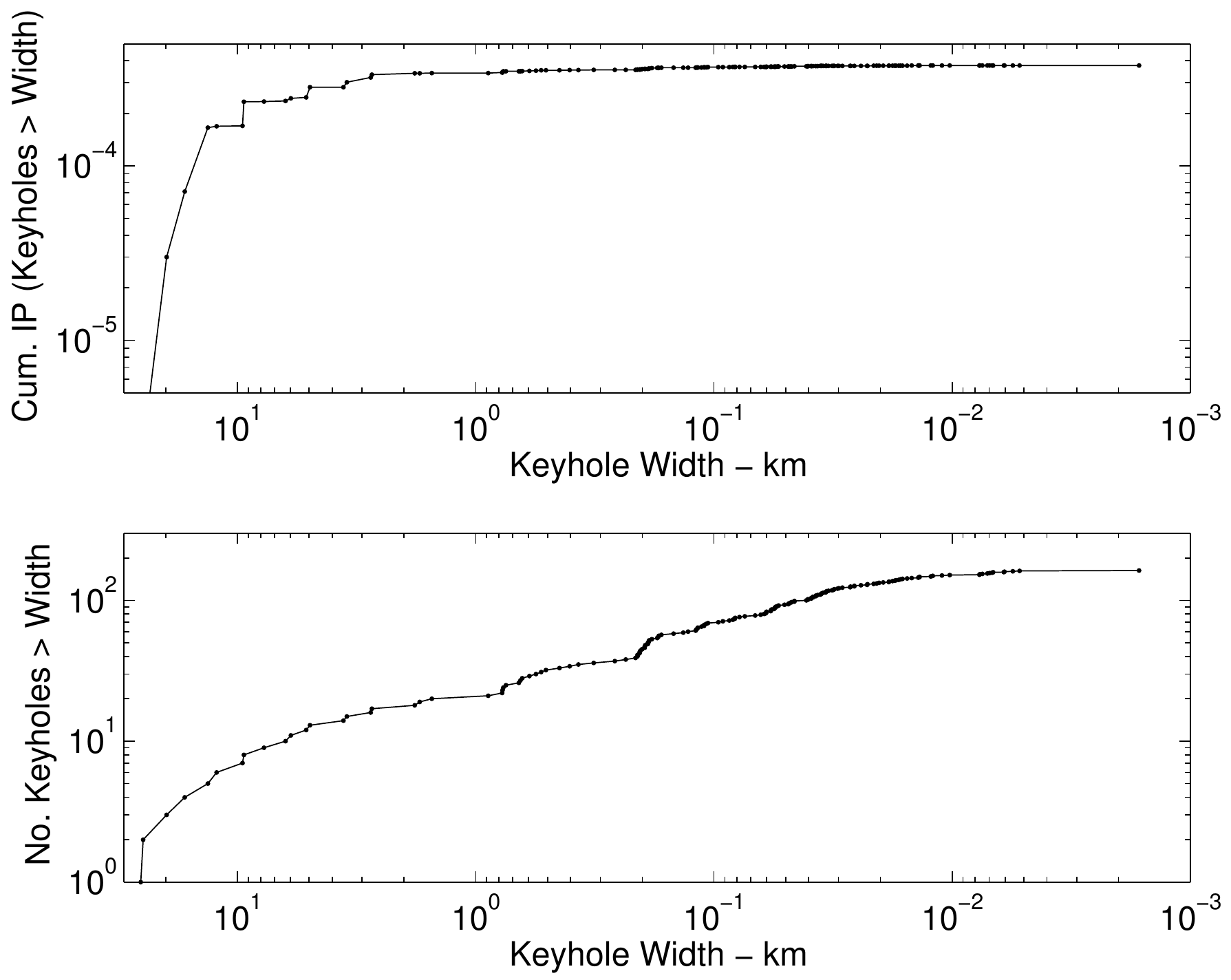}
\caption{Cumulative impact probability (upper) and cumulative number
of keyholes found (lower), each as a function of diminishing keyhole
width. The cumulative impact probability over all potential impacts is
$3.7\times 10^{-4}$.}
\label{fig:cumplot}
\end{center}
\end{figure}

Figure~\ref{fig:cumplot} shows the dependence of the number of keyholes and the cumulative IP on the minimum keyhole width.  Although the number of keyholes increases with decreasing minimum width, the cumulative IP is essentially captured already by only the largest $\sim 10\%$ of keyholes, i.e., those with width $\gtrsim 1$ km.

Post-2135 Earth encounters correspond to resonant returns
\citep{opik_valsecchi}. Table~\ref{tab:resret} describes the main
features of the resonant returns corresponding to an IP $> 10^{-5}$.

It is important to assess the reliability of our results. On one hand,
the keyholes are essentially a geometric factor that does not depend
on the modeling of Bennu's orbit. On the other hand, the PDF on
the 2135 $b$-plane can be strongly affected by the dynamical model and
the statistical treatment applied to the
observations. Table~\ref{tab:models} reports the 2135 $b$-plane
coordinates as a function of the different configurations of the
dynamical model and different settings for the removal of outliers
from the observational data set. It is worth pointing out that
neglecting the Earth relativistic term produces a large error
comparable to a $3\sigma$ shift in the orbital solution. In contrast, the contribution of solar radiation pressure is rather
small. As already discussed in Sec.~\ref{sec:yarko}, this can be
explained by the fact that the action of solar radiation pressure is
aliased with the solar gravitational acceleration, and neglecting solar radiation pressure in the model is therefore
compensated when fitting the orbital solution to the observations. The
different Yarkovsky models give $\zeta_{2135}$ predictions within
several thousand kilometers of each other.  Interestingly, the shift due to the different astrometric outlier
treatment is comparable to the one due to the relativistic term of the
Earth and much larger than any shift due to the other dynamical
configurations. Table~\ref{tab:sbgm} shows the effect of removing each of the 24
perturbing asteroids included in the dynamical model. Ceres, Pallas,
and Vesta give the largest contributions. Among smaller perturbers,
Hebe and Iris turn out to be the most important.

We used OrbFit (see Sec.~\ref{sec:num_val}) to cross-check the keyhole locations and widths, the PDF of Fig.~\ref{fig:keyhole}, and the sensitivity to the different configurations of the dynamical model.  We found good overall agreement with only one noticeable difference related to the PDF: while the PDF shapes are similar, the peaks are separated by about 40000 km in $\zeta_{2135}$. This difference is related to the 0.2$\sigma$ shift in the nominal solution (see Sec.~\ref{sec:num_val}) and is in part due to the fact that OrbFit presently uses JPL's DE405 planetary ephemeris rather than DE424, which is used in our analysis.

\begin{table}[t]
  \caption{Keyholes in 2135 $b$-plane and associated impact 
    probabilities, JPL solution 76.\label{tab:keyhole}}
\small
\begin{center}
\begin{tabular}{ccrcc|ccrcc}
\hline
Year & $\zeta_{2135}$ & $\zeta$-width & Impact & Palermo & 
Year & $\zeta_{2135}$ & $\zeta$-width & Impact & Palermo \\
     & (km)           &    (km)       & Prob.  & Scale   & 
     & (km)           &    (km)       & Prob.  & Scale   \\
\hline
2192 &  92981 &  0.17 &           $1.04\times 10^{-10}$  & $-8.27$ & 2195 & 343783 &  0.11 &           $4.15\times 10^{-07}$  & $-4.67$ \\
2187 &  93280 &  0.32 &           $2.01\times 10^{-10}$  & $-7.97$ & 2181 & 347699 &  0.78 &           $2.86\times 10^{-06}$  & $-3.80$ \\
2198 &  99162 &  0.25 &           $3.33\times 10^{-10}$  & $-7.78$ & 2175 & 349704 &  1.80 &           $6.43\times 10^{-06}$  & $-3.43$ \\
2187 & 105731 &  0.12 &           $3.44\times 10^{-10}$  & $-7.74$ & 2186 & 367729 &  0.32 &           $8.12\times 10^{-07}$  & $-4.36$ \\
2182 & 105945 &  3.59 &           $1.08\times 10^{-08}$  & $-6.23$ & \bf{2175} & \bf{368877} & \bf{16.67} &   $\mathbf{4.13\times 10^{-05}}$ & $
\mathbf{-2.63}$ \\
2197 & 148127 &  0.19 &           $3.08\times 10^{-08}$  & $-5.81$ & 2181 & 370007 &  0.56 &           $1.35\times 10^{-06}$  & $-4.13$ \\
2182 & 153125 &  6.29 &           $1.42\times 10^{-06}$  & $-4.11$ & 2193 & 391326 &  0.60 &           $8.52\times 10^{-07}$  & $-4.36$ \\
2187 & 153481 &  0.21 &           $4.94\times 10^{-08}$  & $-5.58$ & 2187 & 392308 &  5.97 &           $8.34\times 10^{-06}$  & $-3.35$ \\
2192 & 161203 &  1.53 &           $5.83\times 10^{-07}$  & $-4.52$ & 2198 & 393200 &  0.11 &           $1.47\times 10^{-07}$  & $-5.13$ \\
2193 & 162322 &  0.15 &           $6.04\times 10^{-08}$  & $-5.51$ & 2181 & 405893 &  0.64 &           $6.27\times 10^{-07}$  & $-4.46$ \\
2182 & 171554 &  5.15 &           $3.55\times 10^{-06}$  & $-3.71$ & 2187 & 429187 &  0.51 &           $2.43\times 10^{-07}$  & $-4.89$ \\
2187 & 171842 &  0.17 &           $1.19\times 10^{-07}$  & $-5.20$ & 2181 & 430234 &  0.20 &           $9.41\times 10^{-08}$  & $-5.28$ \\
2187 & 174357 &  0.66 &           $5.25\times 10^{-07}$  & $-4.55$ & 2176 & 449272 & 12.23 &           $3.07\times 10^{-06}$  & $-3.76$ \\
2192 & 174935 &  0.40 &           $3.31\times 10^{-07}$  & $-4.76$ & 2193 & 459392 &  0.12 &           $2.09\times 10^{-08}$  & $-5.97$ \\
2194 & 177766 &  0.17 &           $1.57\times 10^{-07}$  & $-5.09$ & 2188 & 475391 &  7.74 &           $7.99\times 10^{-07}$  & $-4.37$ \\
2199 & 185793 &  0.19 &           $2.56\times 10^{-07}$  & $-4.89$ & 2194 & 479545 &  9.54 &           $8.50\times 10^{-07}$  & $-4.36$ \\
\bf{2193} & \bf{186415} & \bf{19.84} &   $\mathbf{2.75\times 10^{-05}}$ & $\mathbf{-2.85}$ & 2194 & 486578 & 25.51 &           $1.77\times 
10^{-06}$  & $-4.04$ \\
2198 & 186996 &  0.21 &           $2.97\times 10^{-07}$  & $-4.83$ & 2194 & 511097 & 24.90 &           $6.94\times 10^{-07}$  & $-4.45$ \\
2199 & 191883 &  0.20 &           $3.53\times 10^{-07}$  & $-4.75$ & 2188 & 517881 &  0.75 &           $1.61\times 10^{-08}$  & $-6.07$ \\
2198 & 208524 &  0.53 &           $1.53\times 10^{-06}$  & $-4.11$ & 2193 & 527510 &  0.89 &           $1.30\times 10^{-08}$  & $-6.17$ \\
2190 & 211276 &  0.11 &           $3.50\times 10^{-07}$  & $-4.74$ & 2181 & 528840 &  0.65 &           $9.06\times 10^{-09}$  & $-6.30$ \\
2199 & 235014 &  0.17 &           $8.69\times 10^{-07}$  & $-4.36$ & 2193 & 534125 &  0.20 &           $2.21\times 10^{-09}$  & $-6.94$ \\
2192 & 272823 &  0.12 &           $8.39\times 10^{-07}$  & $-4.36$ & 2193 & 534310 &  0.19 &           $2.13\times 10^{-09}$  & $-6.96$ \\
2198 & 273950 &  0.11 &           $7.79\times 10^{-07}$  & $-4.41$ & 2198 & 534527 &  0.45 &           $4.96\times 10^{-09}$  & $-6.60$ \\
2186 & 274243 &  0.21 &           $1.48\times 10^{-06}$  & $-4.10$ & 2198 & 536801 &  0.13 &           $1.30\times 10^{-09}$  & $-7.18$ \\
2191 & 275096 &  0.18 &           $1.29\times 10^{-06}$  & $-4.17$ & 2198 & 536823 &  0.21 &           $2.08\times 10^{-09}$  & $-6.98$ \\
2191 & 275569 &  0.77 &           $5.44\times 10^{-06}$  & $-3.55$ & 2196 & 537259 &  0.12 &           $1.19\times 10^{-09}$  & $-7.22$ \\
2191 & 276541 &  0.20 &           $1.42\times 10^{-06}$  & $-4.13$ & 2189 & 537324 &  0.11 &           $1.09\times 10^{-09}$  & $-7.24$ \\
\bf{2185} & \bf{278479} &  \bf{2.76} &   $\mathbf{1.96\times 10^{-05}}$ & $\mathbf{-2.98}$ & 2199 & 539780 &  0.23 &           $2.11\times 10^{-09}$  
& $-6.98$ \\
\bf{2196} & \bf{279590} &  \bf{4.96} &   $\mathbf{3.52\times 10^{-05}}$ & $-2.75$ & 2185 & 541921 &  0.77 &           $6.37\times 10^{-09}$  & 
$-6.46$ \\
\bf{2196} & \bf{281070} & \bf{13.32} &   $\mathbf{9.45\times 10^{-05}}$ & $\mathbf{-2.32}$ & 2185 & 542397 &  1.72 &           $1.39\times 10^{-08}$  
& $-6.12$ \\
\bf{2185} & \bf{295318} &  \bf{9.42} &   $\mathbf{6.33\times 10^{-05}}$ & $\mathbf{-2.47}$ & 2185 & 543753 &  0.37 &           $2.84\times 10^{-09}$  
& $-6.81$ \\
2190 & 302940 &  0.20 &           $1.24\times 10^{-06}$  & $-4.18$ & 2197 & 545747 &  0.26 &           $1.84\times 10^{-09}$  & $-7.03$ \\
\bf{2180} & \bf{316352} &  \bf{3.48} &   $\mathbf{1.95\times 10^{-05}}$ & $\mathbf{-2.96}$ & 2190 & 558169 & 11.57 &           $4.86\times 10^{-08}$  
& $-5.59$ \\
2191 & 316680 &  0.19 &           $1.05\times 10^{-06}$  & $-4.26$ & 2196 & 561278 & 12.31 &           $4.54\times 10^{-08}$  & $-5.64$ \\
2186 & 326599 &  0.19 &           $9.38\times 10^{-07}$  & $-4.30$ & 2196 & 565865 & 20.20 &           $6.12\times 10^{-08}$  & $-5.51$ \\
\bf{2180} & \bf{339506} &  \bf{2.73} &   $\mathbf{1.14\times 10^{-05}}$ & $\mathbf{-3.20}$ & 2190 & 618399 & 17.45 &           $4.95\times 10^{-09}$  
& $-6.59$ \\
2191 & 339838 &  0.13 &           $5.60\times 10^{-07}$  & $-4.53$ & 2184 & 636299 & 54.23 &           $6.51\times 10^{-09}$  & $-6.45$ \\
2190 & 343541 &  0.17 &           $6.74\times 10^{-07}$  & $-4.45$ & 2184 & 651571 &  4.35 &           $2.45\times 10^{-10}$  & $-7.88$ \\
\hline
\end{tabular}
\end{center}
{\em Note}: Impact probabilities $> 10^{-5}$ are highlighted in
bold.
\end{table}

\begin{table}[t]
\caption{Resonances associated with the eight potential impacts with
impact probability $>10^{-5}$.\label{tab:resret}} 
\small
\begin{center}
\begin{tabular}{cccccc}
\hline
  Year &  $\zeta_{2135}$ & Post-2135 Period & Resonance & Res. Period & $\Delta P$ \\
       &  (km)           & (yr)             &           & (yr)        & (yr)       \\
\hline
 2193  &  186415  &   1.2342   & 58 yr/47 rev &  1.2340 &  $+0.0002$  \\
 2185  &  278479  &   1.2215   & 50 yr/41 rev &  1.2195 &  $+0.0020$  \\
 2196  &  279590  &   1.2213   & 61 yr/50 rev &  1.2200 &  $+0.0013$  \\
 2196  &  281070  &   1.2211   & 61 yr/50 rev &  1.2200 &  $+0.0011$  \\
 2185  &  295318  &   1.2194   & 50 yr/41 rev &  1.2195 &  $-0.0001$  \\
 2180  &  316352  &   1.2169   & 45 yr/37 rev &  1.2162 &  $+0.0007$  \\
 2180  &  339506  &   1.2144   & 45 yr/37 rev &  1.2162 &  $-0.0018$  \\
 2175  &  368877  &   1.2116   & 40 yr/33 rev &  1.2121 &  $-0.0005$  \\
\hline
\end{tabular}
\end{center}
\end{table}

\subsection{Statistical Close Approach Frequency}
We now want to characterize the Earth encounter history for Bennu's orbital geometry. The first step is to understand the statistical properties of Earth encounters during a node crossing cycle (see Fig.~\ref{fig:moid}). For this we generated a dense sampling of 20,000 Virtual Asteroids on the  Solution 87 orbit (Table~\ref{tab:nominal}), but with a uniform sampling of the mean anomaly from the full range, $0$ to $2\pi$, to randomize the node crossing trajectory. For each VA we recorded all of the close approaches within 0.015 au during JPL's DE424 ephemerides time interval, i.e., from year $-3000$ to year 3000, which contains only one node crossing cycle. 

\begin{figure}[t]
\begin{center}
\includegraphics[width=5in]{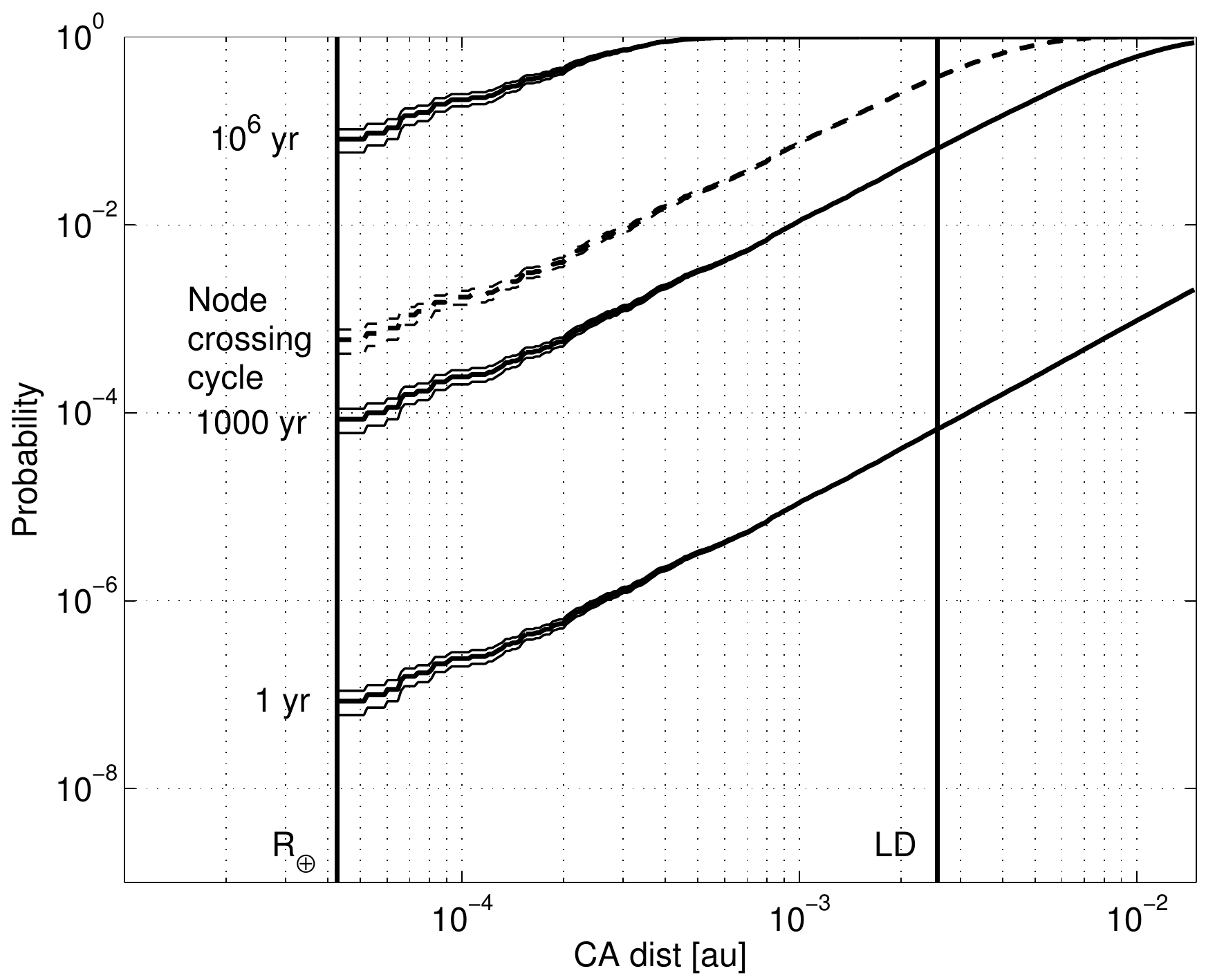}
\caption{Probability (with corresponding $1\sigma$ error bars) of having at least one close approach (CA) within a given distance for different time intervals. The dashed curve is for a node crossing cycle, while solid line are for 1 yr, 1000 yr, and 1,000,000 yr. The Earth radius ($R_\oplus$) and lunar distance (LD) are marked by vertical lines.}
\label{fig:prob_ca}
\end{center}
\end{figure}

We modeled the number of Earth approaches within a given distance in a given time frame as a Poisson random variable. We estimated the Poisson parameter $\lambda$ by averaging over the trajectories of the VAs. Figure~\ref{fig:prob_ca} shows the probability of having at least one close approach within a given geocentric distance during a node crossing cycle (dashed line). For instance, during each node crossing cycle we have 38\% probability of a close approach within the lunar distance and a $6\times 10^{-4}$ probability of an impact. This is consistent with our predictions for the next node crossing taking place around 2200, for which we have similar probabilities of impact and sub-lunar distance encounters.

To analyze the long-term history we need to account for the secular evolution of Bennu's orbit. As reported by NEODyS\footnote{http://newton.dm.unipi.it/neodys/index.php?pc=1.1.6\&n=bennu}, Bennu's perihelion precession period is 28100 yr and each precession period contains four node crossings. For a given time interval we can compute the expected number of node crossings and suitably scale the probability of an encounter within a given distance during a single node crossing cycle. The solid lines in Fig.~\ref{fig:prob_ca} show the probability of at least one Earth encounter within a given distance for time intervals of 1 yr, 1000 yr, and 1,000,000 yr. For example, in a 1000 yr time interval the probability of a close encounter within a lunar distance is 7\% while the probability of an impact is $9 \times 10^{-5}$. This indicates that for Bennu's current orbital configuration the mean Earth impact interval is $\sim10$ Myr. Note that the precession period assumed here is for the nominal orbit of Table~\ref{tab:nominal}, while the precession period does change due to planetary interactions. For instance, the nominal semimajor axis increases and the uncertainty grows after the 2135 encounter, causing the post-2135 precession period to be in the range 28900--33400 years. \citet{delbo_rq36} analyze the orbital evolution of Bennu on a much longer time frame than a single node crossing and find that the median lifetime of Bennu could be $\sim 34$ Myr, but their study allowed for substantial orbital evolution to take place, while our results are valid for the present-day, un-evolved orbit. 

\section{OSIRIS-REx Science}\label{sec:orex}

Continued study of Bennu's trajectory is a significant element of the OSIRIS-REx science investigation. In particular, the characterization of the Yarkovsky effect is planned to be conducted on two tracks. On one track, Earth-based radio tracking of the spacecraft and optical navigation images of the asteroid from the spacecraft will be used to derive high-precision asteroid position measurements. These position updates will afford refined estimates of the nongravitational accelerations that the asteroid experiences. On the other track, science observations by the OSIRIS-REx spacecraft will allow the development of a complete thermophysical model of the asteroid, yielding a precise estimate of the thermal recoil acceleration, as well as direct and reflected solar radiation pressure acting on the body. A comparison of the acceleration profile from these two independent approaches will provide significant insight into the quality of current thermophysical models, and, for example, the extent to which surface roughness affects the net thermal recoil acceleration \citep{rozitis_green_yarko}.

But first the OSIRIS-REx must rendezvous with Bennu, and knowledge of the asteroid position is required for accurate navigation of the spacecraft during the initial encounter. Our current prediction calls for Radial-Transverse-Normal (RTN) position uncertainties of (3.3, 3.8, 6.9) km on 2018-Sep-10, during the planned OSIRIS-REx rendezvous. These are formal 1-sigma error bars, and may not account for some unmodelled systematic effects, although we are not aware of any that are significant. In any case, such low uncertainties  suggest that asteroid ephemeris errors will not be a significant complicating factor during the OSIRIS-REx rendezvous with Bennu. 

To characterize the Bennu ephemeris improvement provided by the OSIRIS-REx mission, we simulate 8 post-rendezvous, pseudo-range points from the geocenter to the asteroid center of mass. The simulated measurements are placed at monthly intervals from 2018-Dec-01 to 2019-Jul-01, and they assume an a priori uncertainty of 0.1 $\mu$s in time delay, which translates to 15 m in range. The trajectory constraints from the OSIRIS-REx radio science effort are likely to be somewhat better than assumed for this study. Table~\ref{tab:orex2} lists the uncertainties obtained before and after the inclusion of these simulated OSIRIS-REx radio science data. We find that the uncertainty in the transverse nongravitational acceleration parameter $A_T$, and by extension the uncertainty in the mean $da/dt$, drops by roughly a factor 6--7, bringing the precision to better than 0.1\%. 

The OSIRIS-REx radio science observations will not only refine the Yarkovsky acceleration acting on the asteroid, but also enable significantly improved future predictions. Table~\ref{tab:orex2} reveals that our current predictions call for position uncertainties of a few km at the end of proximity operations on 2020-Jan-04, which could be reduced to under 100 m with the simulated mission data. The associated velocity uncertainties are of order 1 mm/s with current information, but could fall by a factor 50 or more with the OSIRIS-REx data.

Similarly, we find that the OSIRIS-REx radio science data could narrow the $\zeta_{2135}$ uncertainty region on the 2135 $b$-plane by a factor $\sim 60$. This would be similar to the reduction in uncertainty seen between the \citet{milani_etal_rq36} paper and the present paper. The implication is that the hazard assessment will be dramatically altered by the OSIRIS-REx radio science effort. The self-similar nature of the keyholes on the 2135 $b$-plane suggest that the cumulative probability is likely to remain around $10^{-4}$, although if the nominal $\zeta_{2135}$ prediction does not change appreciably the cluster of relatively wide keyholes near the current nominal (Fig.~\ref{fig:keyhole}) could lead to a cumulative probability of impact in excess of $10^{-2}$.

Besides providing direct radio science position measurements of the asteroid, OSIRIS-REx will refine and test other aspects of the Bennu ephemeris problem.  The mission objectives include 
\begin{itemize}
\item a search for outgassing and the incorporation of any activity into force models,
\item direct measurement of the asteroid mass, providing ground truth for the mass determination technique presented here,
\item precision radiometry of both reflected and thermally emitted radiation with high spatial resolution, providing ground truth for the thermal accelerations presented in this paper, and
\item analysis of the returned sample, which will provide a direct measurement of the thermal, dielectric, and bulk density of the asteroid surface.
\end{itemize}

\begin{table}[t]
\caption{Formal uncertainties with and without simulated OSIRIS-REx pseudo-range measurements as described in the text.\label{tab:orex2}} 
\small
\begin{center}
\begin{tabular}{lcc}
\hline
                            & \multicolumn{2}{c}{Uncertainties} \\
                            &  Reference  & With sim. obs. \\
\hline
 $A_T$ ($10^{-16}$ au/d)    & 2.52 & 0.38\\[1mm]
 \multicolumn{3}{l}{2020-Jan-04 Position (km)}\\
 \hspace{5mm} Radial        & 1.539& 0.008\\ 
 \hspace{5mm} Transverse    & 0.855& 0.033\\ 
 \hspace{5mm} Normal        & 2.461& 0.098\\[1mm]
 \multicolumn{3}{l}{2020-Jan-04 Velocity (mm/s)}\\
 \hspace{5mm} Radial        & 0.662& 0.005\\ 
 \hspace{5mm} Transverse    & 0.400& 0.002\\ 
 \hspace{5mm} Normal        & 1.388& 0.029\\[1mm]
 $\zeta_{2135}$ (km)        & 58000& 1000 \\
 \hline
\end{tabular}
\end{center}
\end{table}

\section{Discussion and Conclusions}

Understanding of an asteroid's physical properties becomes essential whenever the Yarkovsky effect or other nongravitational accelerations are a crucial aspect of the orbit estimation problem. Radar astrometry of asteroids can provide surprising and important constraints, not only on an asteroid's orbit, but also on its physical properties. In the case of Bennu, this information has immense value for space mission designers. We have seen that the availability of well-distributed radar astrometry over time spans of order  a decade can constrain asteroid orbits to the extent that precise estimates of the Yarkovsky effect can be derived. When coupled with thermal inertia information derived from other sources, such as the Spitzer Space Telescope, important parameters such as mass, bulk density and porosity can be derived. Combining Yarkovsky detections with thermal inertia measurements to infer the asteroid mass can be implemented on other near-Earth asteroids, including potential space mission targets. This technique is the focus of ongoing work. Indeed, Bennu clearly demonstrates that even weak radar detections can have considerable science value, raising the imperative to aggressively pursue every available radar ranging opportunity for potential Yarkovsky candidates.

Our bulk density estimate for Bennu implies a primitive body with high porosity of $40\pm10$\%. The implication is that Bennu must be comprised of a gravitationally bound aggregate of rubble, a conclusion that is reinforced by its shape, which is spheroidal with an equatorial bulge consistent with the downslope movement and accumulation of loose material at the potential minimum found at the equator \citep{nolan_etal_2013}. This bodes well for the OSIRIS-REx sample collection effort, which requires loose material at the surface for a successful sample collection, although nothing in this study constrains the size distribution of the surface material.

The statistical encounter frequency with Earth (Fig.~\ref{fig:prob_ca}) can be used to understand the rate of encounters that could alter the shape and spin state of a body through tidal interactions \citep[e.g.,][]{walsh_richardson_2008, nesvorny_etal_2010}. \citet{scheeres_apophis_2005} have shown that tidal interactions at a distance of ~6 Earth radii can appreciably alter the spin state of 99942 Apophis. More distant encounters could still excite the spin state enough to induce seismic activity, leading to a periodic resurfacing of the asteroid that may have implications for interpretation of Bennu samples returned by OSIRIS-REx. 

We have seen that the current levels of uncertainty in Bennu's orbit are low enough that unprecedented levels of accuracy are required in the dynamical model that governs the trajectory. For example, the relativistic perturbation of planetary gravity fields, in particular that of Earth, must be incorporated to obtain reliable results. The future addition of OSIRIS-REx radio science data will again decrease the orbital uncertainties by 1--2 orders of magnitude, which will likely require even finer scale refinements to our dynamical model than used here. 
However, difficulties in understanding the proper statistical treatment of asteroid optical astrometry, and in particular the identification of statistical outliers, will likely remain a dominant source of uncertainty that is not properly captured by a posteriori covariance analysis. 

Thus our findings for the post-OSIRIS-REx orbital uncertainties of Bennu may be illusory. The finding that $\zeta_{2135}$ uncertainties may be reduced as low as 1000 km with the OSIRIS-REx radio science data assumes that our Yarkovsky model, including the asteroid spin state, holds through 2135. Thus a host of model refinements may be necessary to properly characterize the trajectory out to 2135. Notwithstanding the next radar observation opportunity in January 2037, we may reach an uncertainty limit that prevents us from improving predictions any further until models can improve or the prediction interval is significantly reduced. As an example, the post-2135 predictability will markedly improve after the 0.005 au Earth close approach in 2060, and it is reasonable to expect that at least some potential impacts will persist until that time.

\section*{Acknowledgments}

\noindent We are grateful to Giovanni F. Gronchi (Univ. Pisa) for his assistance in calculating Bennu's orbital precession period and its variability.

\noindent This research was conducted in part at the Jet Propulsion
Laboratory, California Institute of Technology, under a contract with
the National Aeronautics and Space Administration.

\noindent D.F. was supported in part by an appointment to the NASA Postdoctoral Program at the Jet Propulsion Laboratory, California Institute of Technology, administered by Oak Ridge Associated Universities through a contract with NASA. Prior to Sept. 2012 his support was through SpaceDys s.r.l. under a contract with the European Space Agency.

\noindent The work of D.V. was partially supported by the Czech Grant Agency (grant P209-13-01308S).

\noindent The work of B.R. is supported by the UK Science and Technology Facilities Council (STFC).

\noindent The Arecibo Observatory is operated by SRI International under a cooperative agreement with the National Science Foundation (AST-1100968), and in alliance with Ana G. Mendez-Universidad Metropolitana, and the Universities Space Research Association. At the time of the observations used in this paper, the Arecibo Observatory was operated by Cornell University under a cooperative agreement with NSF and with support from NASA.

\noindent The Arecibo Planetary Radar Program is supported by the National Aeronautics and Space Administration under Grant No. NNX12AF24G issued through the Near Earth Object Observations program.

\noindent This research was supported in part by NASA under the Science Mission Directorate Research and Analysis Program.

\bibliography{neo}
\bibliographystyle{apalike}

\appendix
\label{appendix}
\section{Finding the best power law to model the Yarkovsky effect}
We want to find the value of $d$ such that the transverse acceleration
$A_T(r/r_0)^{-d}$, $r_0 =1$ au, provides the best match to the
Yarkovsky acceleration acting on Bennu. We can neglect the
seasonal component of the Yarkovsky effect. In fact, the diurnal
component is usually the dominant one \citep{vok_etal_2000}. Moreover,
the obliquity of Bennu is $175^\circ$ and the seasonal
component vanishes when the spin axis is normal to the orbital plane
\citep{bottke_etal06}. For the diurnal component of the Yarkovsky
effect we have that the transverse acceleration is \citep{vok_1998}
\begin{equation}
a_t = \frac{4(1-A)}{9} \phi(r) f(\Theta) \cos(\gamma)\ , \ f(\Theta) =
\frac{0.5 \Theta}{1 + \Theta + 0.5\Theta^2}
\end{equation}
where $A$ is the Bond albedo, $\phi$ the standard radiation force
factor at heliocentric distance $r$, $\Theta$ the thermal parameter,
and $\gamma$ the obliquity.

The dependence on $r$ is contained in $\phi(r)\propto r^{-2}$ and
$f(\Theta)$. As a matter of fact, $\Theta$ depends on the subsolar
temperature $T_\star$, and $T_\star$ depends on $r$:
\begin{equation}
  \Theta = \frac{\Gamma}{\epsilon\sigma T_\star^3}\sqrt{\frac{2\pi}{P}}\
  ,\ T_\star = \sqrt[4]{\frac{(1-A)G_S}{\sigma\epsilon}\left(\frac{1\mbox{ au}}{r}\right)^2}
\end{equation}
where $\Gamma$ is the thermal inertia, $\epsilon$ the emissivity,
$\sigma$ the Stefan-Boltzmann constant, $G_S=1365\, \rm{W}/\rm{m}^2$ is the solar constant,
and $P$ the rotation period. Thus, $\Theta \propto r^{3/2}$.

We want to approximate $f(\Theta)$ with a power law $(r/r_0
)^\psi$:
\begin{equation}
f(\Theta) \simeq c(r/r_0)^\psi \Longrightarrow \log f(\Theta) \simeq \log c
+ \psi \log (r/r_0)\ .
\end{equation}
By differentiating with respect to $r$ we find:
\begin{equation}
  \psi = r\ \frac{\partial \log f(\Theta)}{\partial r}\bigg|_{r = r_0}\ .
\end{equation}
From the chain rule we obtain
\begin{equation}
  \frac{\partial \log f(\Theta)}{\partial r} =
  \frac{1}{f(\Theta)}\frac{\partial f(\Theta)}{\partial
    \Theta} \frac{\partial \Theta}{\partial r}
\end{equation}
Evaluating this equation for $r=r_0$ and using the physical quantities
as in Table~\ref{tab:rq36_props} yield $\psi \simeq
-0.24$. Therefore, $d\simeq 2.24$, which is in good agreement with
the value 2.25 found numerically in Sec.~\ref{sec:yarko}.

\begin{figure}[t]
\begin{center}
\includegraphics[width=5in]{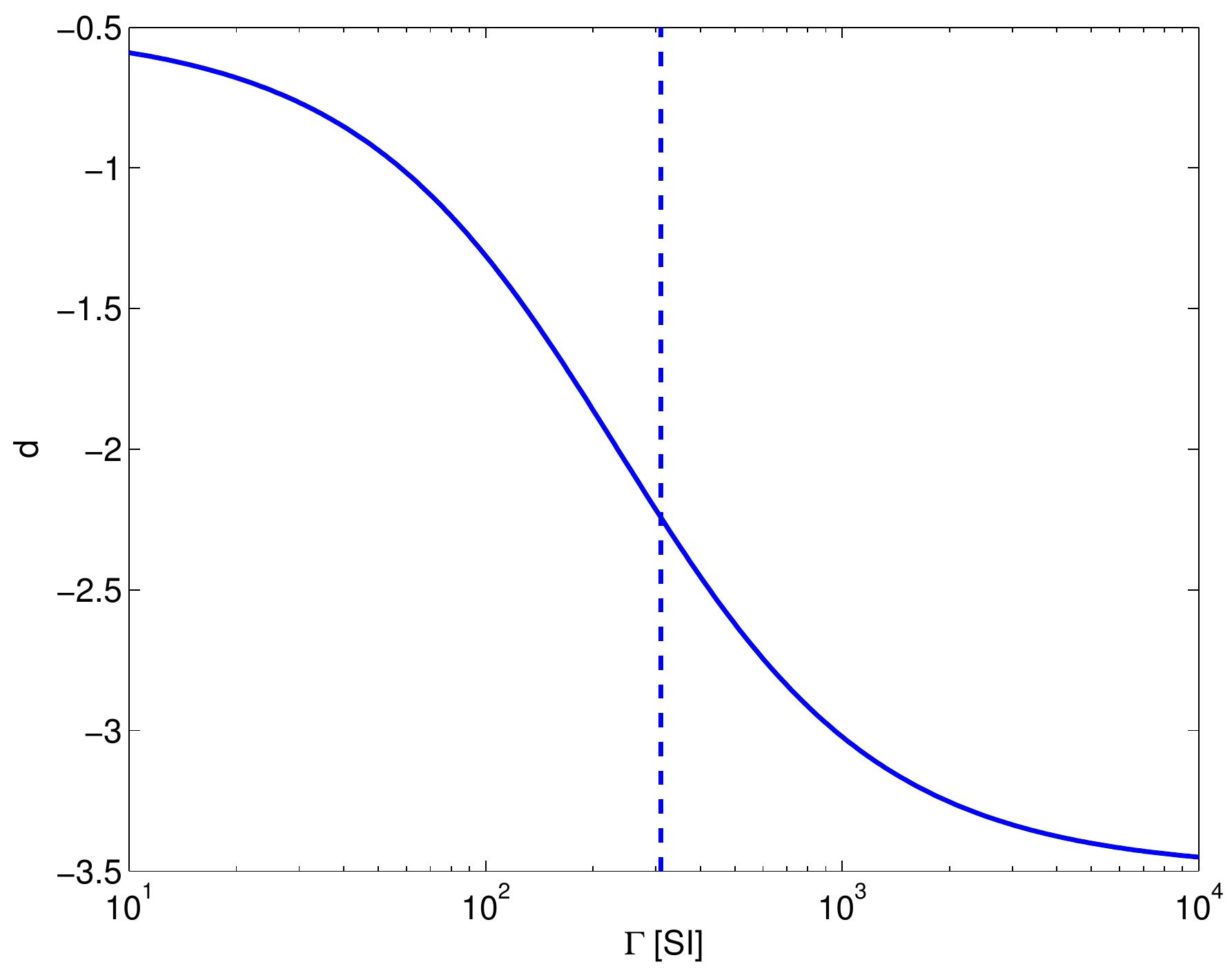}
\caption{Dependency of transverse acceleration exponent $d$ on thermal inertia $\Gamma$ for Bennu.}
\end{center}
\end{figure}


\vspace{3cm}
\noindent \copyright 2014. All rights reserved.

\end{document}